\documentclass[11pt]{article}

\pdfoutput=1
\usepackage{color}
\usepackage{amsmath,amssymb,bm,url}
\usepackage{slashed,relsize}
\usepackage{graphicx}
\usepackage[T1]{fontenc}
\usepackage{cite}
\usepackage{hyperref}
\def\beq{\begin{equation}\displaystyle}
\def\eeq{\end{equation}}
\def\bea{\begin{eqnarray}\displaystyle}
\def\eea{\end{eqnarray}}
\newcommand{\be}{\begin{eqnarray}}
\newcommand{\ee}{\end{eqnarray}}
\definecolor{colorRTD}{rgb}{.2,.2,.7}
\definecolor{colorGG}{rgb}{.3,.5,.7}

\usepackage{float}
\usepackage{subfig}
\usepackage{adjustbox}
\usepackage{multirow}
\usepackage{comment}
\usepackage{placeins}
\usepackage{fancyhdr}

\begin{document}

%\baselineskip=13pt

%\date{\today}
%

\author{Raffaele Tito D'Agnolo$^{1}$,\, Gaia Grosso$^{2,3}$,\, Maurizio Pierini$^{3}$,\,
  Andrea Wulzer$^{2,4,5}$,\,Marco Zanetti$^{2}$\\[2pt]
{\small\emph{$^{1}$ {Institut de Physique Th\'eorique, Universit\'e Paris Saclay, CEA, F-91191 Gif-sur-Yvette, France}}}\\
{\small\emph{$^{2}$Dipartimento di Fisica e Astronomia, Universit\`a di Padova and}}\\
{\small\emph{INFN, Sezione di Padova, via Marzolo 8, I-35131 Padova, Italy}}\\
{\small\emph{$^{3}$CERN, Experimental Physics Department, Geneva, Switzerland}}\\
{\small\emph{$^{4}$CERN, Theoretical Physics Department, Geneva, Switzerland}}\\
{\small\emph{$^{5}$ Theoretical Particle Physics Laboratory (LPTP), Institute of Physics, EPFL, Lausanne, Switzerland}}
}

\title{\bf{Learning Multivariate New Physics}}

\maketitle

\date{}

%%%%%%%%%%%%%%%%%%%%%%%%%%%%%%%%
\abstract{
We discuss a method that employs a multilayer perceptron to detect deviations from a reference model in large multivariate datasets. Our data analysis strategy does not rely on any prior assumption on the nature of the deviation. It is designed to be sensitive to small discrepancies that arise  in datasets dominated by the reference model. The main conceptual building blocks were introduced in Ref.~\cite{DAgnolo:2018cun}. Here we make decisive progress in the algorithm implementation and we demonstrate its applicability to problems in high energy physics. We show that the method is sensitive to  putative new physics signals in di-muon final states at the LHC. We also compare our performances on toy problems with the ones of alternative methods proposed in the literature.}
%%%%%%%%%%%%%%%%%%%%%%%%%%%%%%%%

%%%%%%%%%%%%%%%%%%%%%%%%%%%%%%%%%%%%%%%
%\setcounter{page}{2}
%%%%%%%%%%%%%%%%%%%%%%%%%%%%%%%%%%%%%%%

\thispagestyle{fancy}

\newpage
%%%%%%%%%%%%%%%%%%%%%%%%%%%%%%%%
\section{Introduction}\label{sec:intro}
In the study of the fundamental laws of Nature we face a number of open questions. In the past decades the field of particle physics has produced a set of potential answers that seemed inevitable in their simplicity. The experimental effort inspired by these solutions is now mature and is slowly stripping them of their initial theoretical appeal. As more and more data are collected, the problems that confront us become sharper and harder to solve. We know that the theories that well describe current data are incomplete and should be extended, but our prior beliefs on how the extension should look like and on where to discover it experimentally become less concordant every day. %At this stage, the best course of action is to turn to experiments for answers. Especially to experiments that are already successfully running and exploring aspects of Nature that we did not have access to before. 
In this paper we show how to interrogate experimental data in a new way, going beyond searches targeted at one specific theoretical model. 

We consider the problem of having large multivariate datasets that are seemingly well described by a reference model. Departures from the reference model can be statistically significant, but are caused only by a very small fraction of events. The significance of the discrepancy might stem from the extreme rarity of the discrepant events in the reference model and in this case standard anomaly detection techniques might be employed. Or the discrepancy is due to a small excess (or even a deficit) of events in a region of the space of physical observables that is also populated in the reference model. Our goal is to determine if the experimental dataset does follow the reference model exactly or if it instead contains ``small'' departures as described above. In the latter case, we also want to know in which region of the space of observables the discrepancy is localized. This problem is relevant to Large Hadron Collider (LHC) datasets that are well described by the Standard Model of particle physics (SM) and CMB datasets that are well described by the standard cosmological model $\Lambda$CDM. 

Our focus here will be physics cases relevant to the LHC. Many attempts at generalizing traditional new physics searches based on specific models have already been made in this context, developing what are called ``model independent'' search strategies. As customary in the literature~\cite{Choudalakis:2011qn, Abbott:2000fb, Abbott:2000gx, Aktas:2004pz, Aaron:2008aa, Asadi:2017qon, Aaltonen:2007dg, Aaltonen:2008vt, CMS:2008gya, CMS:2011fra, ATLAS:2017irs, ATLAS:2012qna, ATLAS:2014sxa, Arguelles:2019izp} by model independent analysis we mean an analysis that does not target a specific new physics model. Typically, analysis techniques of this kind assume that a background model is at hand. This could be obtained from simulation or some simulation-based reweighting of a data control region.\footnote{Also in the approach we follow in the present paper, both options are viable since we only need a reference data sample distributed as the SM predicts. It might come from a Monte Carlo simulation or from a control region.}  Recent papers (e.g., Ref.~\cite{Andreassen_2020}) started referring to data-driven background estimates as (background) model-independence, to be accompanied by (signal) model-independence for a truly model-independent approach. We do not employ this latter notion of {\it model independence} in this study, and we keep the estimate of the background predictions, with the associated uncertainties, as a separated aspect (see below) from the one of model-independence. Notice that model-independent analyses (no matter which notion of model-independence is adopted) do rely on the choice of the specific final state (reconstructed variables and acceptance cuts) to which they are applied. This obviously restricts the sensitivity only to new physics models that contribute to the specific final state, and reintroduces some amount of model-dependence. 

Model-independent analyses typically follow the binned histogram technique, in which one selects a set of bins (i.e. search regions) in the space of observables and compares the amount of data observed in each bin with the reference model (i.e., the SM). The main problem with this approach comes from the fact that, as previously emphasized, the data distribution will be identical to the one predicted by the SM in the vast majority of the phase space. The observed countings in almost all bins will thus be in agreement with the SM expectation, but only up to the unavoidable Poisson fluctuations. Poisson fluctuations from non-discrepant bins do contribute to the binned likelihood, and if there are many non-discrepant bins their contribution will overwhelm the contribution the likelihood receives from the few genuinely discrepant bins. This is not an issue in a binned (or unbinned) analysis targeting a specific new physics model, because the bins where no discrepancies are expected do not contribute to the likelihood (which is the ratio between the new physics and the SM Poisson likelihoods). In the model-independent case instead, Poisson fluctuations from non-discrepant bins easily swamp any potential signal of new physics. Typically this can be mitigated only by paying a high price in flexibility \cite{Choudalakis:2011qn, Collins:2018epr}.

In this work we apply a new methodology to the problem, expanding on the ideas presented in Ref.~\cite{DAgnolo:2018cun}. Our technique leverages the progress that the field of machine learning has experienced in the past few years. In particular we exploit the flexibility of neural networks as multidimensional function approximants~\cite{Cybenko1989,KREINOVICH1991381,Hecht-Nielsen:1992:TBN:140639.140643,HORNIK1989359,2016arXiv161004161L,DBLP:journals/corr/PoggioMRML16,DBLP:journals/corr/Bach14,2017arXiv171208688M}. Here we show that this idea addresses the challenge presented above for realistic multidimensional datasets and physically motivated putative signals. 
In particular we consider $\mu^+\mu^-$ production at the LHC and we quantify the sensitivity of our method to a resonance $Z^\prime\rightarrow\mu^+\mu^-$ and to a non-resonant signal induced by a four-fermion contact interaction. 

It should be stressed that the design of the algorithm is purely based on the knowledge of the reference model with the criteria described in Section~\ref{sec:method}. No optimization was performed based on the putative new physics signals, as appropriate for a model-independent search strategy. We always present the sensitivity of our method in comparison with the ``ideal'' sensitivity one might obtain with a standard model-dependent search strategy that is instead optimized for the specific model at hand. We will also discuss how the trained neural network can help identifying the physical origin of the observed discrepancy.

At this stage, we assume that the reference dataset provides a perfect representation of the background distribution in real data. In a typical analysis, this is true within the effect of systematic uncertainties, which are controlled by nuisance parameters. The effect of these nuisance parameters can be accounted for in our method, by a straightforward application of the profile likelihood ratio methodology. The practical implementation of this extension of the method will be the topic of a future publication~\cite{future}. In this study, we ignore this aspect and concentrate on the more pressing issue of generalizing Ref.~\cite{DAgnolo:2018cun} to a multivariate problem. 
This motivates the choice of relatively simple and clean experimental signature: that in fact allows introducing the new method and discussing its strength, knowing that the underlying assumptions (e.g., the possibility of accessing a trustable reference sample with larger statistics than the data sample) will be fulfilled. It is reasonable to expect that extending this method to other final states (e.g., dijet) might imply additional practical problems (e.g., the need of a large reference dataset).

Our results benefit from a crucial methodological advance that we make in this paper compared to Ref.~\cite{DAgnolo:2018cun}. This consists in an algorithmic procedure to select the regularization parameters of the neural network and the network architecture. We take the regularization parameter to be a hard upper bound (weight clipping) on the magnitude of the weights. While admittedly heuristic (even if based on robust results in statistics), we will see that this procedure uniquely selects the weight clipping and it also gives constraints on the viable neural network architectures\footnote{While more standard validation procedures like k-folding could be exploited, in this study we only focus on the approach described in Section \ref{sec:alg}, 
specifically designed to fit the proposed methodology.}. 

Machine learning techniques have recently been introduced to solve problems related to the one discussed above~\cite{Weisser:2016cnc, Blance:2019ibf, Heimel:2018mkt, Cerri:2018anq, Farina:2018fyg, Hajer:2018kqm, DeSimone:2018efk}. In this paper we also directly compare our sensitivity with that of two related ideas presented in the literature. One has the same goal, but is based on a nearest neighbors estimation of probability distributions~\cite{DeSimone:2018efk, Hajer:2018kqm}. The other targets only resonant signals, with the resonant feature occurring in a pre-specified variable, but leverages in a similar way the capability of multilayer perceptrons to identify correlations in multivariate datasets~\cite{Collins:2018epr}. For the comparison we employ simple toy benchmark examples defined in the corresponding publications. We study these examples with our method and compare our performances with the published results. This is a first step towards an exhaustive comparison of the different proposals (that also include ~\cite{Blance:2019ibf, Heimel:2018mkt, Cerri:2018anq, Farina:2018fyg, Hajer:2018kqm, Weisser:2016cnc}), which we consider necessary at this stage given the practical difficulties involved in directly evaluating their respective strengths and weaknesses by just reading published work. 

The paper is organized as follows. In Section~\ref{sec:method}, after a brief review of the basic ideas behind our approach (see Ref.~\cite{DAgnolo:2018cun} for a detailed exposition), we define its detailed implementation. We describe in particular the strategy we adopt to select the neural network architecture and the other hyperparameters. In Section~\ref{sec:otherworks} we compare our performances with Refs.~\cite{DeSimone:2018efk,Hajer:2018kqm,Collins:2018epr} in the context of toy examples. The rest of the paper is devoted to $\mu^+\mu^-$ production at the LHC. First, in Section~\ref{sec:datasets}, we introduce the new physics signals and the details of our simulated datasets. We also describe the dedicated analyses that we use to estimate the ideal sensitivity. In Section~\ref{sec:results} we describe the application of our method and we extensively study its performances. We conclude and outline directions for future work in Section~\ref{sec:conc}.

%%%%%%%%%%%%%%%%%%%%%%%%%%%%%%%
\section{Methodology}\label{sec:method}
Neural networks have already found a plethora of successful applications in high energy physics, including jet physics \cite{deOliveira:2015xxd, Schwartzman:2016jqu, Kagan:2016wnu,	Larkoski:2017jix, Louppe:2017ipp, Shimmin:2017mfk, Baldi:2016fql, Guest:2016iqz, Almeida:2015jua, Barnard:2016qma, Kasieczka:2017nvn, Butter:2017cot, Datta:2017rhs, Datta:2017lxt, Fraser:2018ieu, Andreassen:2018apy, Macaluso:2018tck, ATLAS:2017jiz, CMS-DP-2017-013, ATL-PHYS-PUB-2017-013, ATL-PHYS-PUB-2017-004, CMS-DP-2017-005, ATL-PHYS-PUB-2017-003, ATL-PHYS-PUB-2017-017, CMS-DP-2017-027}, optimized new physics searches~\cite{Baldi:2016fzo, Chang:2017kvc, Cohen:2017exh, Brehmer:2018eca, Brehmer:2018kdj, Brehmer:2018hga, Roxlo:2018adx, Collins:2018epr}, faster detector simulations~\cite{Paganini:2017dwg,deOliveira:2017rwa,Paganini:2017hrr, Freitas:2019hbk, Brehmer:2019xox, Brehmer:2019gmn} and fits to parton distribution functions \cite{Ball:2014uwa}, where they have been  applied successfully for decades~\cite{Forte:2002fg}. In this work we show the power of these techniques in the context of model-independent new physics searches at the LHC, expanding the framework of Ref.~\cite{DAgnolo:2018cun}.

We first choose a set of variables that describe the data, a range for their values and the integrated luminosity of the dataset. This is the only physics choice that we have to make, which defines the ``experiment'' we want to analyze. For instance our input space can consist of the momenta of the two leading muons in events with at least two opposite-sign muons within acceptance. 

Once we have selected an input space of interest we generate a large reference sample that represents the SM (or, ``reference model'') prediction. This simulated dataset has much larger statistics than the actual experimental data, we denote with $\mathcal{N}_R$ the number of events in this sample, while the expected number of events is $N(R)\ll \mathcal{N}_R$. We also generate $N_{\rm toy}$ toy datasets that again follow the SM prediction, but have the same statistics as the actual experimental dataset. Namely, they contain a variable number of events, 
%${\mathcal{N}}_{\mathcal{D}}$ events, 
thrown randomly from a Poisson distribution with $N(R)$ expected events. At this point we have prepared the required input for the neural network and we can choose a specific network architecture. 

Our neural networks are fully connected, feedforward regressors, trained to learn a likelihood ratio. The training is carried on with a supervised procedure, taking as input the two datasets described above: the large reference dataset that follows the SM (reference) prediction $\mathcal{R}$ and a smaller dataset that represents the experimental data $\mathcal{D}$. The training datasets are preprocessed. Input variables allowing negative values, as $\eta$, are normalized subtracting their mean and dividing by their standard deviation. The other variables, like $p_T$, are simply divided by their mean. The loss $L$ used to train the network is 
\be
L[f(\,\cdot\,,{\mathbf{w}})]=\frac{N({\textrm{R}})}{{\mathcal{N}}_{\mathcal{R}}}\sum\limits_{x\in{\mathcal{R}}}(e^{f(x;{{{\mathbf{w}}}})}-1)-\sum\limits_{x\in {\mathcal{D}}}f(x;{{{\mathbf{w}}}})\, . \label{eq:loss}
\ee
Here $x$ is an element of the input space (for example $5$ numbers describing the two muons $p_T$, rapidity and azimuthal angular difference) and $f$ is the output of the network as a function of the free parameters $\mathbf{w}$ (weights and biases) of the network. The values of these parameters after training will be denoted as $\mathbf{\widehat{w}}$ in what follows.

The neural network defines a composite hypothesis for the distribution (denoted as $n(x| {\bf w})$) of the data, namely
$$
\displaystyle
n(x| {\bf w})\equiv e^{f(x;{{{\mathbf{w}}}})}n(x|{\rm{R}})\,,
$$
where $n(x|{\rm{R}})$ is the distribution (see Table~\ref{tab:notation} for a summary of our notation) in the reference hypothesis. Our search strategy is constructed as an hypothesis test between the simple hypothesis $n(x|{\rm{R}})$ and the composite (depending on the free parameters ${\mathbf{w}}$) alternative hypothesis  $n(x| {\bf w})$. The loss is constructed~\cite{DAgnolo:2018cun} to reproduce the maximum log-likelihood ratio (Neyman--Pearson) test statistic $t(\mathcal{D})$ for composite alternative hypothesis~\cite{10.2307/91247}. Namely it is such that
\be\label{MLL}
\underset{\{{\mathbf{w}}\}}{\rm Min}\; L = - \underset{\{{\mathbf{w}}\}}{\rm Max}\;\left\{ \log\left[\frac{e^{-N({{\mathbf{{w}}}})}}{e^{-N({\rm{R}})}}\prod\limits_{x\in {\mathcal{D}}}\frac{n(x|{{\mathbf{{w}}}})}{n(x|{\rm{R}})}\right]\right\}= -\frac{t(\mathcal{D})}{2}\,,
\ee
when ${\mathcal{N}}_{\mathcal{R}}$ is much larger than the number of events in ${\mathcal{D}}$. The minimum of the loss at the end of training (or, more precisely, ${t(\mathcal{D})}$) is thus employed as the statistic of our hypothesis test. Notice that after training, the output of the network is an estimate of the log-ratio of the data distribution over the reference distribution: $f(x;\mathbf{\widehat w})\simeq\log \left[n(x|\mathbf{\widehat{w}})/n(x|{\rm{R}})\right]$, as a function of the input variables $x$. It can thus be used, in case of tension between the data and the reference model, to identify the most discrepant regions of the phase space.

\begin{table}[!b]
\caption{Summary of notation introduced and employed in Sections~\ref{sec:method} and \ref{sec:datasets}.}
\begin{center}
\begin{tabular}{|c|c|}
\hline
& {\bf Distributions}\\ \hline
$n(x|{\rm{R}})$ & Distribution of the variable $x$ in the reference model ${\rm{R}}$ \\ \hline
$n(x|{\rm{NP}})$ & Distribution in the new physics model ${\rm{NP}}$ (signal plus background) \\ \hline
%$n(x|{\rm{T}})$ & True distribution of $x$ \\ \hline
$n(x|\widehat {\bf w})$ & Distribution of $x$ estimated by the Neural Network (NN)\\ \hline \hline 
& {\bf Events}\\
 \hline
$N({\rm{R}})$ & Number of expected events in the reference model ${\rm{R}}$ \\ \hline
%%% ADDED
${\mathcal{N}}_{\mathcal{R}}$ & Number of events in the reference dataset \\ \hline
%${\mathcal{N}}_{\mathcal{D}}$ & Number of events in the Data sample $\mathcal{D}$ \\ \hline
%%%%%%
$N(\widehat {\bf w})$ & Number of expected events estimated  by the NN\\ \hline \hline 
& {\bf Test Statistic}\\
 \hline
$t({\mathcal{D}})$ & Test statistic computed by the NN on the Data sample $\mathcal{D}$ \\ \hline
$t_{\rm{id}}({\mathcal{D}})$ & Ideal test statistic (requires prior knowledge of the signal) \\ \hline
$P(t|{\rm{R}})$ & Probability distribution of the test statistic $t$ in the reference model \\ \hline
%$P(t|{\rm NP})$ & Probability distribution of the test statistic $t$ in the new physics model NP \\ \hline \hline
& {\bf Normalization}\\ \hline
$\int \hspace{-2pt}n(x) dx =N$ & $n(x)$: Events distribution \\ \hline
$\int \hspace{-2pt}P(x) dx =1$ & $P(x)$: Probability distribution \\ \hline
\end{tabular}
\end{center}
\label{tab:notation}
\end{table}%
%%%%%%%%%%%%%%%%%%%%%%%%%%%%%%%%%%

%The network lands on the values $\mathbf{\widehat{w}}$ after testing a number of likelihood functions characterized by different values of the parameters: $\mathbf{{w}^\prime}, \mathbf{{w}^{\prime\prime}}, ...$. This is the sense in which we are performing a model-independent search. We have an alternative hypothesis (which makes our procedure well defined from the point of view of hypothesis testing), but the alternative hypothesis is a large ensemble of possible likelihood functions, limited only by the flexibility of the network. 

%We do not know explicitly $n(x|{\rm{R}})$ due to the intricacies of detector simulation. Obtaining a numerical approximation of it by multidimensional histograms from simulation is not feasible in a large number of dimensions. The network learns directly the log-ratio:
%$f(x|\mathbf{\widehat{w}})=\log \left[n(x|\mathbf{\widehat{w}})/n(x|{\rm{R}})\right]$
%$f(x|\mathbf{\widehat{w}})=\log \left[n(x|{\cal D})/n(x|{\rm{R}})\right]$
%by minimizing the loss function
%\be
%L[f(\,\cdot\,,{\mathbf{w}})]=\frac{N({\textrm{R}})}{{\mathcal{N}}_{\mathcal{R}}}\sum\limits_{x\in{\mathcal{R}}}(e^{f(x;{{{\mathbf{w}}}})}-1)-\sum\limits_{x\in {\mathcal{D}}}f(x;{{{\mathbf{w}}}})\, . \label{eq:loss}
%\ee
Armed with the input datasets, the network and loss described above we can analyze the data. The procedure is rather straightforward.
%We first give as input to the network the experimental data and the large reference sample.
%%%%%%%
The neural network is trained using the reference sample and a data sample collected by the experiment. The loss at the end of training produces a single value $t_{\rm obs}$ for the test statistic.
%This produces a single value $t_{\rm obs}$ for the test statistic.
We then train the network again using, instead of the experimental data, the $N_{\rm toy}$ synthetic datasets distributed according to the reference hypothesis, previously described. This gives us $N_{\rm toy}$ values of the test statistic $t$ that populate the distribution of the test statistic in the reference model hypothesis: $P(t|{\rm R})$. Comparing $t_{\rm{obs}}$ with $P(t|{\rm{R}})$ tells us if our dataset is consistent with the reference model. More precisely we can compute a global\footnote{We stress the fact that our p-values are global. i.e., we do account for the look-elsewhere effect in the specific analysis at hand. On the other hand, there is a residual trial factor, induced by repeating the procedure on multiple final states. This trial factor is difficult to quantify, and analogous to the usually neglected trial factor of the global search effort by an LHC experiment consisting of hundreds of searches in different final states. We do not discuss it further here.} $p$-value as
\be
p=\int_{t_{\rm obs}}^\infty dt P(t| {\rm{R}})\,.
\ee
%where P(t| {\rm{R}}) is estimated from the $N_{\rm toy}$ toy datasets. 
We also define a corresponding $Z$ score as
\be
Z(p)=\Phi^{-1}(1-p)\; ,
\ee
where $\Phi^{-1}$ is the quantile of a Normal distribution with zero mean and unitary variance, so that $Z$ is conveniently expressed as a number of $\sigma$'s.
The presence of a new physics signal in the experimental dataset would manifest itself as a large value of Z. 
%If this is observed one can go back and analyze the output of the network trained with experimental data:  $f(x;\mathbf{\widehat w})\simeq \log[n(x|\mathbf{\widehat{w}})/n(x|{\rm{R}})]$. This is a fully transparent physical quantity and can be used for systematic cross checks. This procedure is amenable to the inclusion of systematic uncertainties. We will comment more on this important aspect in the Conclusions.

The discussion above concisely lays out our data analysis strategy (see Ref.~\cite{DAgnolo:2018cun} for a more complete exposition), to be put in place once the neural network architecture and the other hyperparameters have been selected. We now describe the criteria and the algorithmic procedure by which this selection is made, which constitute the major methodological advance of the present paper.

%\section{The network architecture}
\subsection{Hyperparameters Selection}\label{sec:alg} 
Our goal is to design an effectively model independent search, so the construction of our method must not assume to know anything specific about the signal that we are looking for.
%%%%%%%%%%%%%%%%%%  NEW SENTENCE   %%%%%%%%%%%%%%%%%%%
%so our method does not assume to know anything specific about the signal that we are looking for. Regardless of this, a single analysis based on our method carries a residual model dependence, related to the choice of the input features. This choice is intrinsically related to the LHC data taking strategy: every dataset at hand comes from a trigger selection that implies some selection on a set of objects (muons, jets, etc.). The most general feature choice one can make is then to define the space of the input features by using the four-momenta of these objects. Doing so on an ensemble of different final states, originating from different trigger selections, one would recover model independence as a whole.
%%%%%%%%%%%%%%%%%%  NEW SENTENCE   %%%%%%%%%%%%%%%%%%%
Our selection strategy is thus purely based on the  reference model (SM) prediction, and relies on two general criteria.
%Clearly this implies that our choice of hyperparameters is not the optimal one if one wants to detect a single specific signal.

The first criterion that we adopt is flexibility. Namely we would like the neural network to have as many parameters as possible, free to vary in the largest possible range, in order to be sensitive to the largest possible variety of new physics. 

This has to be balanced against a second criterion, based on the following observation. Our method is mathematically equivalent to the Maximum Likelihood hypothesis test strategy where the set of alternative hypotheses is defined by the neural network. Hence we can rely on the classical results by Wilks and Wald~\cite{Wilks:1938dza, Wald1943} (see also \cite{Cowan:2010js} for a more recent exposition) according to which the maximum log-likelihood ratio test statistics is distributed in the Asymptotic Limit as a $\chi^2$ with a number of degrees of freedom equal to the number of free parameters in the alternative probability model. From here we conclude that the distribution of our test statistic on reference-model toy datasets (i.e., $P(t|{\rm R})$) approaches in the Asymptotic Limit a $\chi^2$ with a number of degrees of freedom given by the number of parameters of the neural network. Clearly we should not expect this result to hold for a finite dataset. However if it does apply, namely if the distribution does resemble the $\chi^2$, we can conclude heuristically that the dataset is sufficiently abundant for the network that is being fitted. If instead the distribution violates the asymptotic formula, it means that the test statistic is sensitive to low-statistics regions of the dataset that are subject to large and uncontrolled fluctuations. We can define this behavior as ``overfitting'' in our context and restrict ourselves to hyperparameters configurations for which a good compatibility of $P(t|{\rm R})$ with the appropriate $\chi^2$ distribution is observed. We will see that combining the two criteria of flexibility and of $\chi^2$-compatibility dramatically restricts the space of viable options.  

\begin{figure}[!t]
\begin{center}
\includegraphics[width=0.9\textwidth]{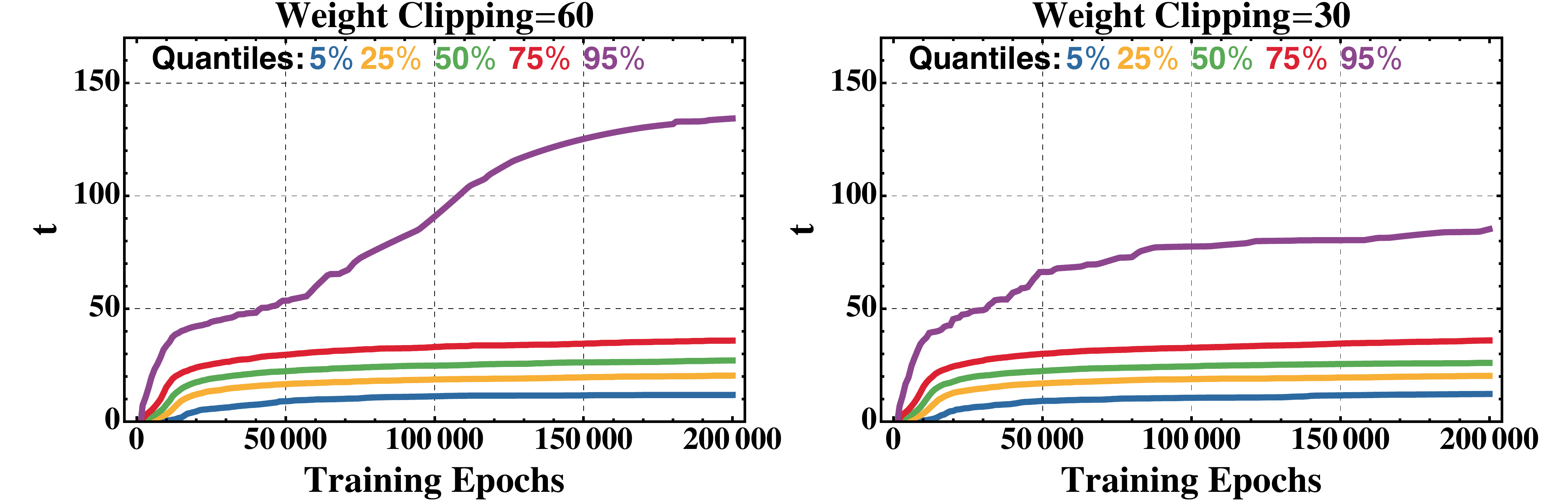} \\ \mbox{} \\
\includegraphics[width=0.9\textwidth]{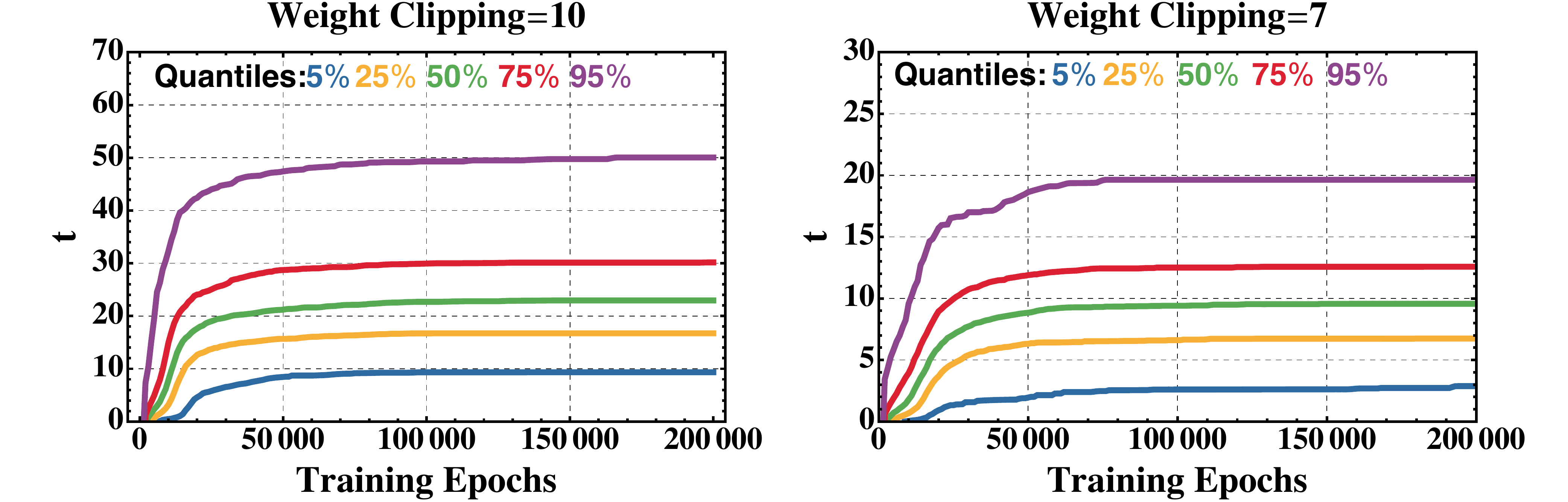}
%    $\begin{array}{ccc}
%    \subfloat[Weight clipping 60]{\includegraphics[height=5.6cm, width=5.cm]{Figures/WC_1}}&
%    \subfloat[Weight clipping 30]{\includegraphics[height=5.6cm, width=5.cm]{Figures/WC_2}}&
%    \subfloat[Weight clipping 10]{\includegraphics[height=5.6cm, width=5.cm]{Figures/WC_3}}\\
%    \end{array}$
    \caption{Quantiles of the test statistic distribution vs training epochs for different choices of Weight Clipping. The quantiles are obtained from 1000 toy experiments and are a plotted for a 1D example discussed in the text. The architecture of the network is fixed at 1-3-1.}\label{fig:outliers}
\end{center}
\end{figure}

In order to illustrate how the optimization strategy works in practice we first need to specify our framework. We restrict ourselves to fully-connected neural networks with logistic sigmoid activation functions in the inner layers. The architecture is characterized by the dimensionality of the input of each of the ``L'' layers, i.e. by a set of integers $a_0$-$a_1$-$\ldots$-$a_{\rm\sc{L-1}}$, plus the output dimensionality that is fixed to $a_{\rm\sc{L}}=1$ in our case. So for example a 1-3-1 ($L=2$) network acts on a one-dimensional feature space ($a_0=1$), has one inner layer with three neurons  ($a_1=3$) and one-dimensional output  ($a_2=1$). In this notation the total number of parameters (weights and biases) in the network is
\beq\label{pareq}
N_{\textrm{par}}(\vec{a})=\sum_{n=1}^{{\rm\sc{L}}} a_{n}(a_{n-1}+1)\,.
\eeq
We regularize the network by imposing an upper bound (Weight Clipping) on the absolute value of each weight. In the following we capitalize Weight Clipping when referring to this specific use of the parameter (i.e., an upper bound on each individual weight). The minimization of the loss function in eq.~(\ref{MLL}) is performed using ADAM~\cite{Kingma:2014vow} as implemented in {\tt Keras}~\cite{Keras} (with the {\tt TensorFlow}~\cite{TensorFlow} backend) with parameters fixed to $\beta_1=0.9$, $\beta_2=0.99$, $\epsilon=10^{-7}$ and initial learning set to $10^{-3}$. The batch size is always fixed to cover the full training sample. The hyperparameters we want to determine are thus the number of layers and of neurons in each layer (i.e., the architecture of the network), the Weight Clipping parameter and the number of training epochs.

The first step of the optimization procedure consists in choosing an initial network architecture. This can be done heuristically by considering the dimension of the input space and the number of events in the datasets of interest. Here we consider for illustration a one-dimensional slice (specifically, the momentum of the leading lepton in the $x$ direction) of the SM di-muon dataset described in Section~\ref{sec:datasets} with a relatively low expected number of data events $N({\rm{R}})=2000$. The number of events in the reference sample is ${\cal{N}}_{\cal{R}}=20000$. A small $1$-$3$-$1$ network is a reasonable starting point in this case. According to the flexibility criterion, the Weight Clipping parameter should be taken as large as possible in order to maximize the expressive power of the network. However if we take it very large training does not converge even after hundreds of thousands of training epochs. This is not acceptable because reaching the absolute minimum of the loss function as in eq.~(\ref{MLL}) is conceptually essential for our strategy. We observe this behaviour in the upper left corner of Figure~\ref{fig:outliers}, where we plot the upper quantiles of $P(t|R)$ as a function of training rounds. The phenomenon is avoided by lowering the Weight Clipping below a certain threshold ${\rm{W}}_{\rm{max}}$, which we find to be ${\rm{W}}_{\rm{max}}\simeq30$ in the case at hand as shown in the figure. 

\begin{figure}[!t]
\begin{center}
\includegraphics[width=0.5\textwidth]{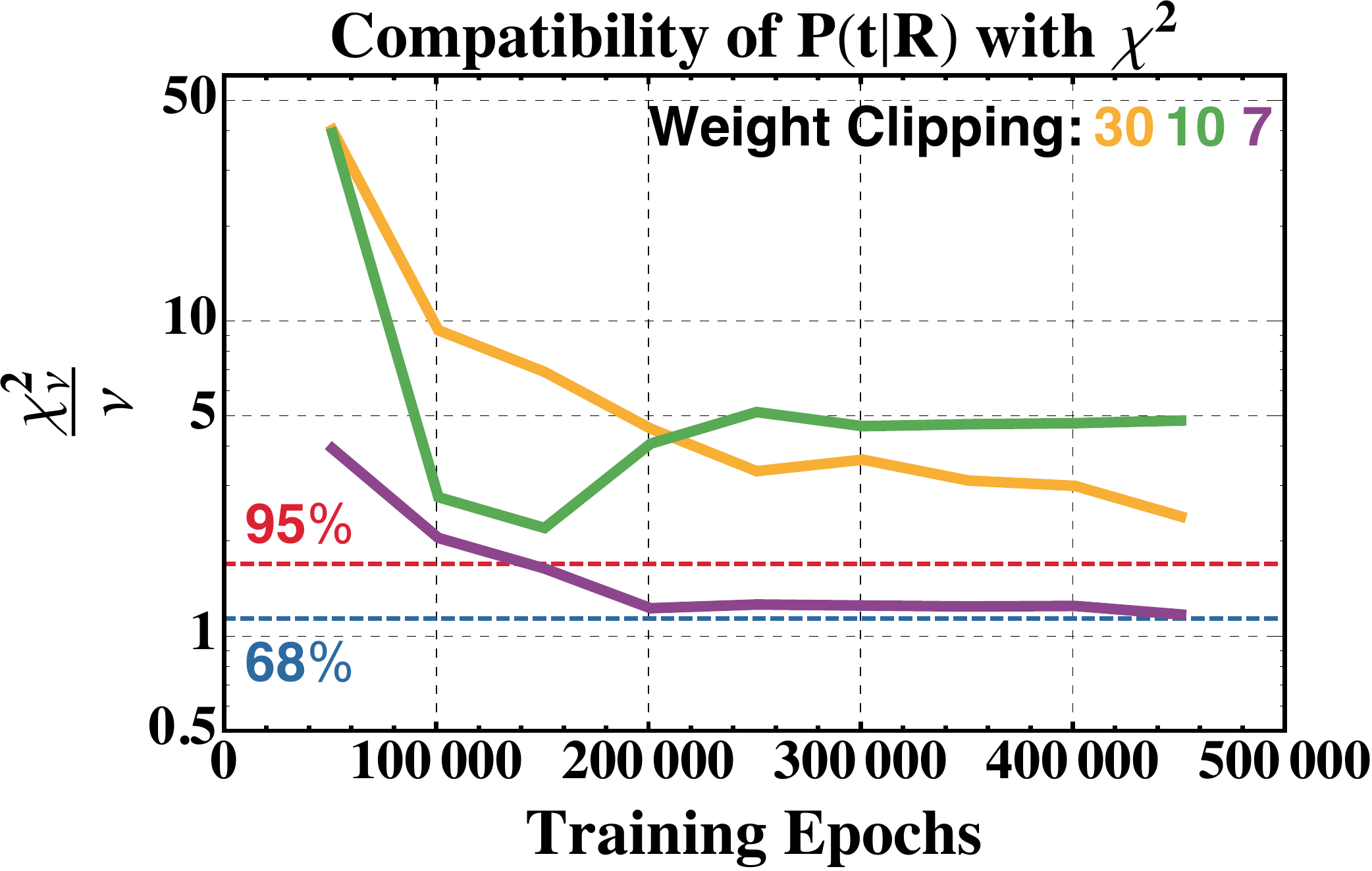}\;\;\;\;
\includegraphics[width=0.45\textwidth]{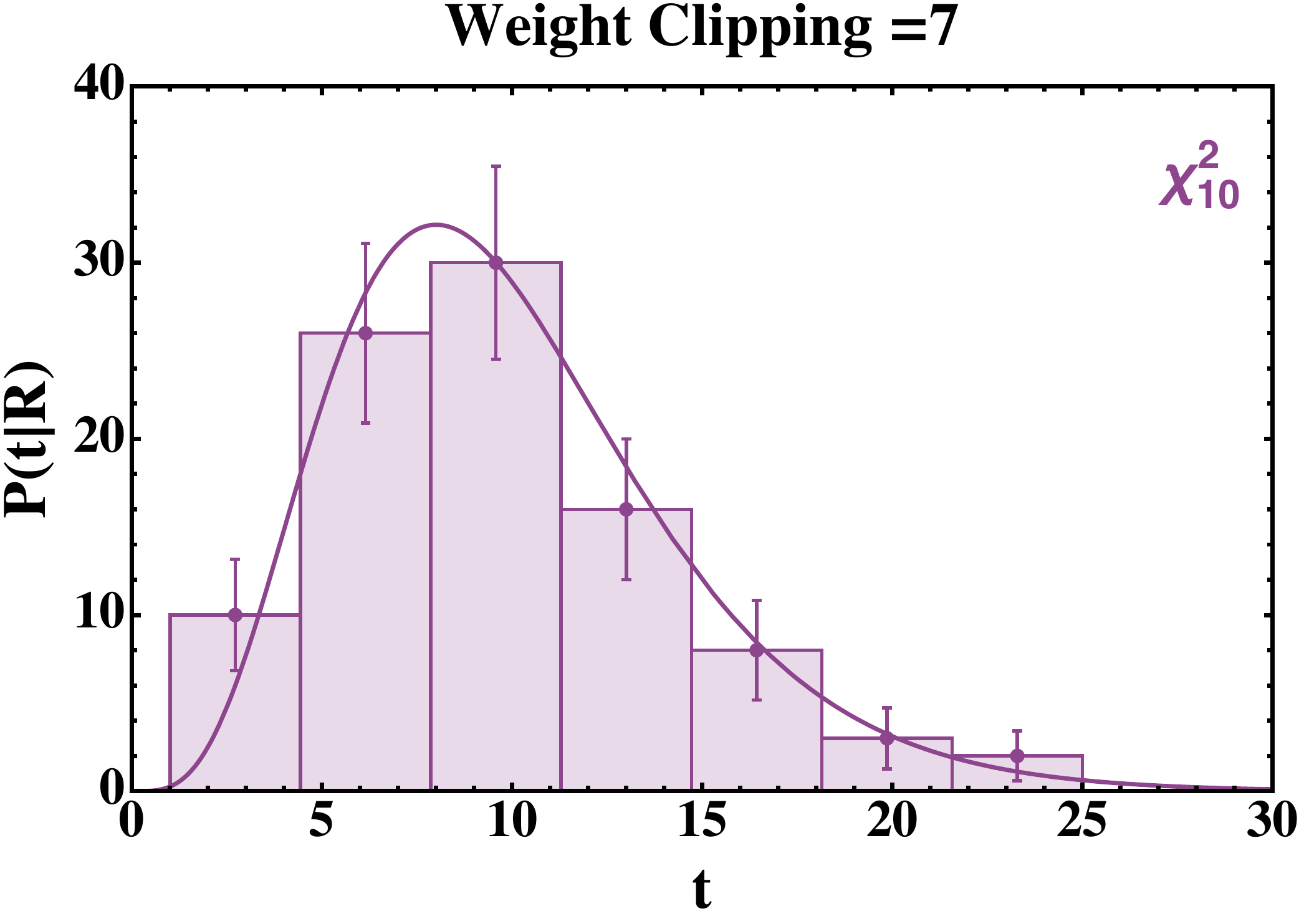}
    \caption{Left: compatibility of the test statistic distribution in the reference hypothesis with a $\chi^2$ distribution with ${N_{\textrm{par}}}=10$ degrees of freedom (1-3-1 network). The plot was made using 100 toy experiments and a 1D example discussed in the text. Note that the $\chi^2_\nu$ on the y-axis measures the compatibility between the two distribution and is not related with the $\chi^2_{N_{\textrm{par}}}$ that approximates reference model distribution of $t$. Right: the test statistic distribution for Weight Clipping set to $7$, compared with the $\chi^2_{10}$. }\label{fig:wtuning}
\end{center}
\end{figure}

The test statistic distribution $P(t|R)$ can now be compared with the $\chi^2_{N_{\textrm{par}}}$ distribution, with a number of degrees of freedom equal to the number of parameters of the neural network as in eq.~(\ref{pareq}). We have $N_{\textrm{par}}=10$ for the 1-3-1 network. The left panel of Figure~\ref{fig:wtuning} displays the evolution with the training rounds of the $\chi^2$-compatibility, defined as a simple Pearson's~$\chi^2$ test statistic on the $P(t|R)$ distribution sampled with $1000$ toy experiments. We see that requiring an acceptable level of $\chi^2$-compatibility further restricts the allowed range for the Weight Clipping parameter. The maximum Weight Clipping for which compatibility is found is $7$ in the case at hand. Since the Weight Clipping should be as large as possible to maximize flexibility, this is the value to be selected.

%%%%%%%%%%%%%%%%%%%%%%%%%%%%%%%%
%%%%%%%%%%%%%%%%%%%%%%%%%%%%%%%%

In summary, the strategy we adopt to select the Weight Clipping parameter is the following:
\begin{enumerate}
\item Starting from a large Weight Clipping, decrease it until the evolution of the 95\% quantiles of $P(t|R)$ achieve a plateau as a function of training epochs.
\item In the range of Weight Clippings below ${\rm{W}}_{\rm{max}}$ where the the plateau is reached, choose the largest Weight Clipping value that gives a good compatibility %\footnote{Here we determine compatibility by building a $\chi^2$ between a binned $P(t|R)$ distribution (15 bins) from 1000 SM toys and the relevant $\chi^2$ distribution.} 
between $P(t|R)$ and a $\chi^2$ distribution whose degrees of freedom are equal the total number of trainable parameters in the network, as shown in Figure~\ref{fig:wtuning}.
\item The total number of training epochs should also be fixed. To reduce the computational burden of our procedure this is chosen as the minimum value for which the evolution of the $\chi^2$-compatibility has reached its plateau.
\end{enumerate}

%%%%%%%%%%%%%%%%%%%%%%%%%%%%%%%%
\begin{figure}[!t]
\begin{center}
\includegraphics[width=0.5\textwidth]{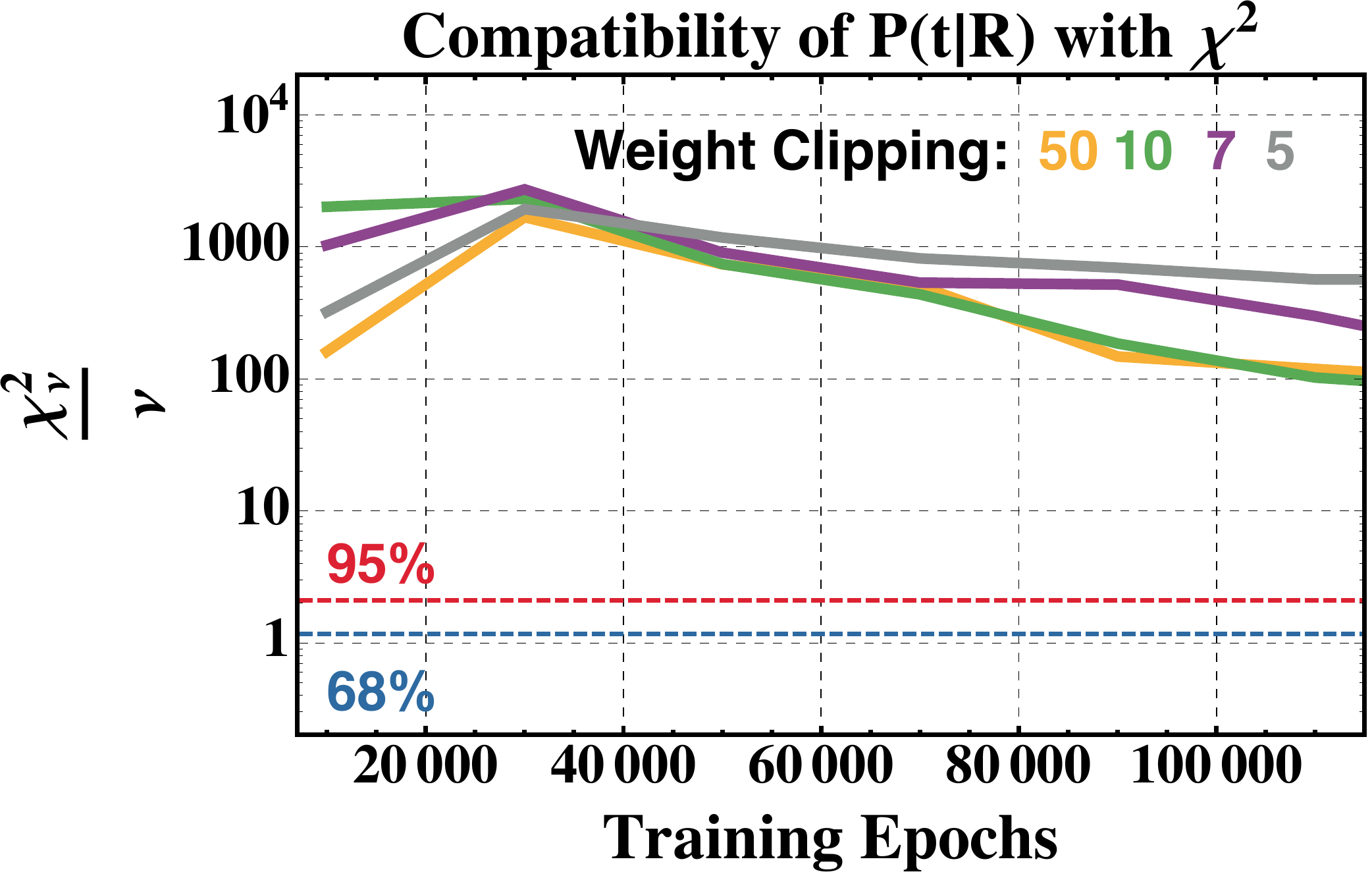}
    \caption{Compatibility of the test statistic distribution in the reference hypothesis with a $\chi^2$ distribution with ${N_{\textrm{par}}}=31$ degrees of freedom (1-10-1 network) as a function of a training rounds and for different choices of the Weight Clipping parameter. A satisfactory level of compatibility is never reached.}\label{fig:wtuningbis}
\end{center}
\end{figure}
%%%%%%%%%%%%%%%%%%%%%%%%%%%%%%%%

We should now explore different neural network architectures. In particular we would like to consider more complex architectures than 1-3-1 to increase the expressive power of the network. Complexity can indeed be increased, but not indefinitely as shown in Figure~\ref{fig:wtuningbis} for a 1-10-1 network. A suitable ${\rm{W}}_{\rm{max}}$ can be identified below which the quantiles of $P(t|R)$ converge, but $P(t|R)$ fails to fulfil the $\chi^2$-compatibility criterion for any choice of the Weight Clipping parameter. The 1-10-1 network should thus be discarded and the optimal (largest) viable network of the 1-N-1 class sits in the range $3\leq {\textrm{N}}<10$. By studying the networks in this range we might uniquely select the architecture and all the other hyperparameters that are suited for the problem at hand.

The behaviour described above for the toy one-dimensional dataset has been confirmed in other cases and it is believed to be of general validity. Namely it is generically true that $\chi^2$-compatibility places an upper bound on the network complexity, leaving us with a finite set of options to explore. On the other hand we cannot claim that the our selection strategy always singles out a unique hyperparameters configuration. Even in our one-dimensional example one might extend the complexity of the network by adding also new layers, obtaining  several viable options with similar number of parameters. Selecting one of these options would require to introduce a strict notion of neural network ``complexity'', to be maximized. Furthermore one might consider departures from the general neural network framework that we are considering. For instance the weight clipping might be imposed on the norm of the weight vector at each layer rather than on each individual weight, and/or a different weight clipping threshold might be imposed on each layer. Even the choice of logistic sigmoid activations and of fully-connected networks might be reconsidered. 

While this aspect should be further studied, it is probably unnecessary to consider this extended space of possibilities. This belief is supported by a number of tests that we performed with different activation functions, training methods and architectures. We find that the performances of our strategy in terms of sensitivity to putative new physics signals depend quite weakly on the detailed implementation of the algorithm. Performances are very similar for all the hyperparameters configurations that are reasonably flexible and obey the $\chi^2$-compatibility criterion. Even slight departures from compatibility typically do not change the sensitivity appreciably. Establishing this fact for a number of putative new physics signals and for several neural network configurations selected with our criteria would justify the choice of restricting to a single configuration. Or, alternatively, would allow to combine the $p$-values obtained from different strategies without loosing sensitivity by the look-elsewhere effect.

Before concluding this section it is worth to point out that the compatibility with the $\chi^2_{N_{\textrm{par}}}$ distribution can be leveraged to compute the $p$-value without generating a large number of toy experiments
\be\label{chi2app}
p=\int_{t_{\rm obs}}^\infty P(t|R) dt \simeq \int_{t_{\rm obs}}^\infty \chi^2_{N_{\textrm{par}}}(t) dt \, .
\ee
This considerably reduces the computational burden of evaluating the global significance. However one should keep in mind that $P(t|R)\simeq\chi^2_{N_{\textrm{par}}}(t)$ is an approximate statement, which we can only test at a statistical level given by the number of toys that we generated. However in cases where generating a sufficient number of toy samples is unfeasible, as for example the high significance models discussed in Section~\ref{sec:otherworks}, we will be obliged to report estimates of the $p$-value obtained with this approximation.

We stress that our hyperparameter selection strategy is not based on an a priori assumed new physics model, hence it is not optimal to detect any specific signal. In particular, the network we select with our method might not be expressive enough to be fully sensitive to complex new physics signals. However we will see this is not a problem for the BSM scenarios studied in this paper.

%The method was designed to identify small signals, as discussed in more detail in the second paragraph of the Introduction. By small signals we mean signals that can be statistically significant, but are caused only by a very small fraction of events, giving an $S+B$ probability distribution which is identical to the background-only distribution in the vast majority of the parameter space. The technique for hyperparameters selection described in this Section relies only on knowledge of the reference model, so it can select less expressive networks than those that would optimally separate signal from background in cases where the $S+B$ probability distribution is very different from the background only distribution. However optimality would not be needed to detect such a large signal. 
%
%On the other hand, cases can be realised with signals differing substantially from the competing background processes; such signals, even if produced with small probability, could be well identified by dedicated and complex enough neural networks. The optimal discrimination could be however be obtained only if the addressed signal is known a priori.  
%In our case the network has to be designed exclusively on the background properties and thus the network resulting from the Weight Clipping procedure may have suboptimal discrimination power.

%%%%%%%%%%%%%%%%%%%%%%%%%%%%%%%
\section{Comparison with Related Work}\label{sec:otherworks}
In the previous section we have introduced all the ingredients needed to implement our data analysis strategy. In this section and in Section~\ref{sec:results} we test its performances on a series of examples. This section is devoted to comparisons with other ideas that have related goals.

Machine learning has recently seen a surge of popularity following the latest developments in deep learning and computer vision. A number of works proposing anomaly detection strategies for LHC datasets has appeared in the literature in the past few years~\cite{Blance:2019ibf, Heimel:2018mkt, Cerri:2018anq, Farina:2018fyg, Hajer:2018kqm, DeSimone:2018efk, Weisser:2016cnc}. This effort is still relatively recent and the field has not fully matured yet.
% new sentence
Ongoing efforts (LHC Olympics~\cite{kasieczka_gregor_2019_3596919,gregor_kasieczka_2019_2629073} and DarkMachines~\cite{darkmachines_community_2020_3685861}) are establishing benchmarks for common comparison.
%The rapidity of its development resulted in a lack of common benchmarks.

We take a step towards making the comparison between different strategies more transparent by testing our methodology on some toy examples present in the literature. We consider three examples: two incarnations of a method that has the same goal, but a very different estimation strategy for the test statistic and a third method that has a narrower scope, but a similar technical approach to the problem, based on multilayer perceptrons trained as classifiers.

The first strategy that we compare with is the nearest neighbors approach of Refs.~\cite{DeSimone:2018efk,Hajer:2018kqm}. This is a truly model-independent approach\footnote{Note that in reality one always needs an alternative hypothesis to obtain a significance. Our use of the term model-independent is explained in the previous section and in Ref.~\cite{DAgnolo:2018cun}.} that aims at reconstructing the true probability distributions for the data and the reference model, using a nearest neighbors technique~\cite{Schilling, Henze, Wang}. A comparison to this method is instructive because it allows us to test a completely different approach to the estimation of the likelihood.

We first study the performance of our algorithm on a two-dimensional example, considered in Ref.~\cite{DeSimone:2018efk}, comprised of events extracted from Normal distributions. The reference model is a two-dimensional Gaussian with mean $\vec \mu=(1,1)$ and covariance matrix $\Sigma = \mathbf{1}_{2\times 2}$. The number of expected events in the reference model is $N({\rm R})=20000$.

We consider two different putative new physics (NP) models:
\begin{itemize}
\item NP$_1$: the data have mean $\vec \mu=(1.12,1.12)$ and covariance matrix $\Sigma = \mathbf{1}_{2\times 2}$. The number of events predicted is the same as in the reference model: $N({\rm NP}_1)=20000$. 
\item NP$_2$: the data have mean $\vec \mu=(1,1)$ and covariance matrix $\Sigma = ((0.95,0.1),(0.1,0.8))$. Again, $N({\rm NP}_2)=20000$. 
\end{itemize}

%%%%%%%%%%%%%%%%%%%%%%%%%%%%%%%%%%%%%%
\begin{figure}[!t]
\begin{center}
\includegraphics[width=0.8\textwidth]{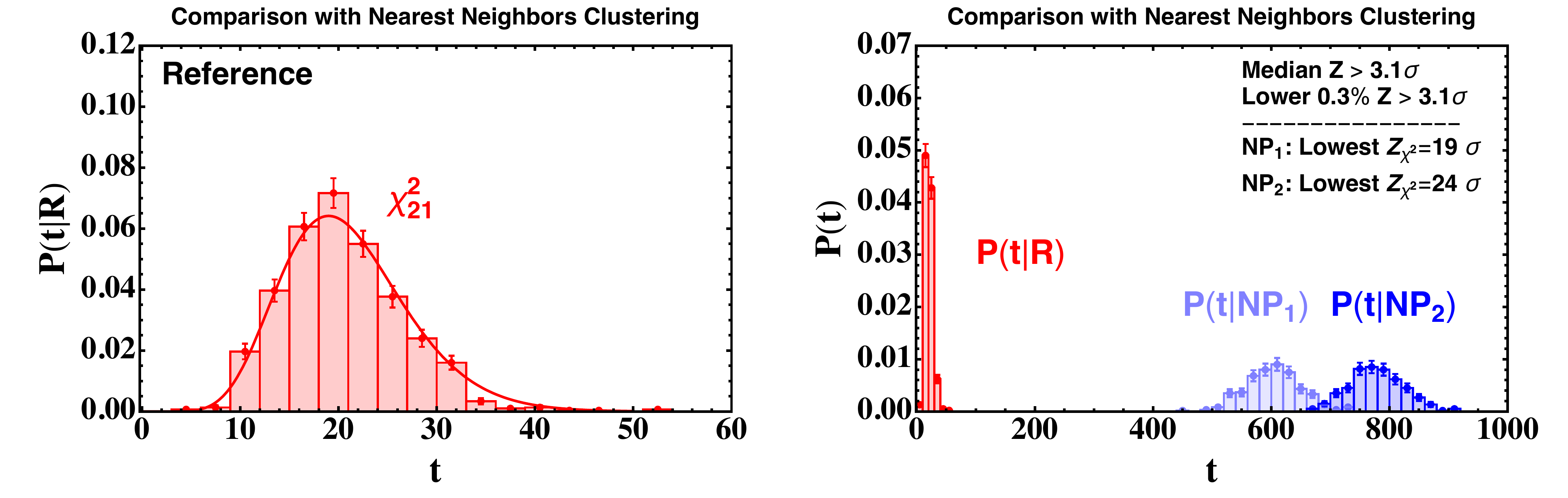}
\includegraphics[width=0.8\textwidth]{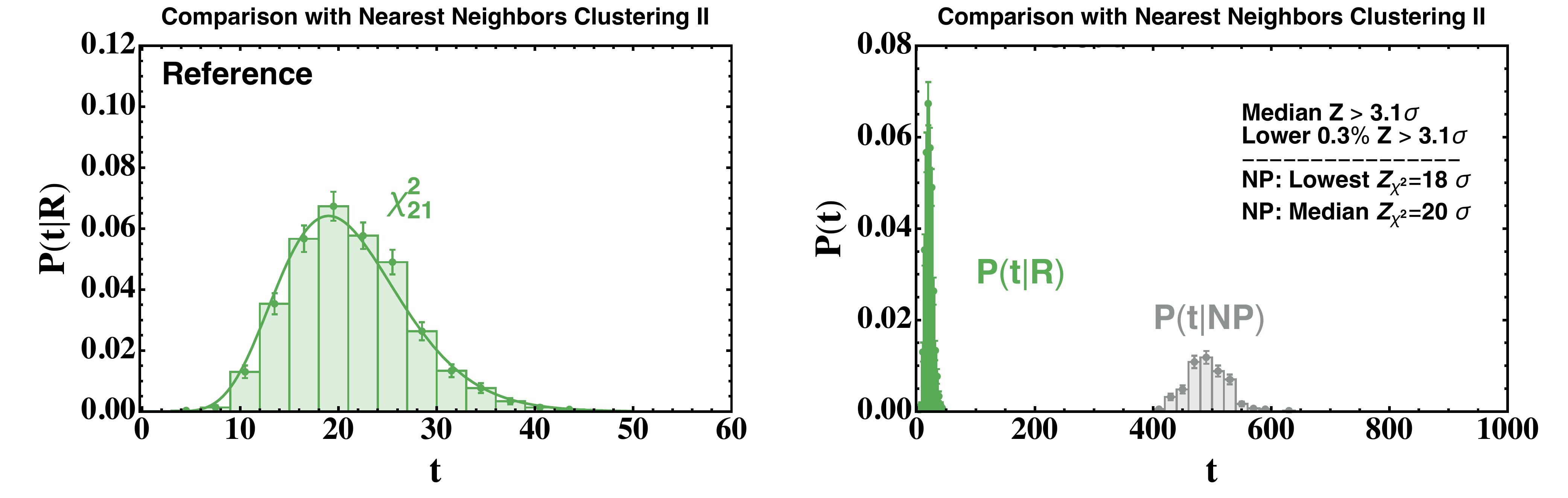}
\caption{{{Top:}} Test statistic distribution for the signal and background models considered in~\cite{DeSimone:2018efk}, obtained with our analysis technique. %The plots represent the Gaussian toy example in 2 dimensions labelled $\mathcal{T}_{G_0}$ (reference), $\mathcal{T}_{G_1}$ (NP$_1$) and $\mathcal{T}_{G_2}$ (NP$_2$). 
The plot displays our sensitivity obtained using the toy samples ($Z$) and the $\chi^2$ approximation of $P(t|R)$ ($Z_\chi^2$). The significance quoted in Ref.~\cite{DeSimone:2018efk} is $2.2\sigma$ for NP$_1$ and $3.5\sigma$ for NP$_2$. {{Bottom:}} Test statistic distribution for the signal and background models considered in~\cite{Hajer:2018kqm}, obtained with our analysis technique.
}\label{fig:DS}
\end{center}
\end{figure}
%%%%%%%%%%%%%%%%%%%%%%%%%%%%%%%%%%%%%%
%%%%%%%%%%%%%%%%%%%%%%%%%%%%%%%%%%%%%%
%\begin{figure}[!t]
%\begin{center}
%\includegraphics[width=0.9\textwidth]{Figures/Nearest_Neighbors}
%\caption{Test statistic distribution for the signal and background models considered in~\cite{Hajer:2018kqm}, obtained with our analysis technique. The plot displays our sensitivity obtained using the toy samples ($Z$) and the $\chi^2$ approximation of $P(t|R)$ ($Z_\chi^2$).}\label{fig:nn2}
%\end{center}
%\end{figure}
%%%%%%%%%%%%%%%%%%%%%%%%%%%%%%%%%%%%%%

The results are shown in Figure~\ref{fig:DS} (top) for a 2-5-1 network with Weight Clipping $1.2$ and 150000 epochs of training. We generated 1000 toy SM samples and 300 data samples distributed according to the new physics hypothesis. The lowest significances that we find, using the $\chi^2$ approximation (\ref{chi2app}) for the reference model test statistic distribution, are $Z_{\chi^2}=19\sigma$ for NP$_1$ and $Z_{\chi^2}=24\sigma$ for NP$_2$. 
%These performances are not surprising given the extremely high ideal significance of this example, way above the usual $5\sigma$ discovery threshold. Due to computational limitations it is hard to attach a precise number to the ideal $Z$-score, but it is above the numbers quoted for our neural network. 
The nearest neighbor approach of~\cite{DeSimone:2018efk} finds $Z=2.2(3.5)\sigma$ for NP$_1$(NP$_2$) for 5 nearest neighbors and 1000 permutations used to estimate the test statistic distribution in the reference hypothesis. 
%\RTD{Spiegare perch\'e il loro approccio fallisce in questi casi}

An alternative implementation of the nearest neighbors approach was proposed in Ref.~\cite{Hajer:2018kqm}. The following two-dimensional problem is considered:
\begin{itemize}
\item Reference model (R): the data have mean $\vec \mu=(0,0)$ and covariance matrix $\Sigma = \mathbf{1}_{2\times 2}$. The number of expected events is $N({\rm R})=10000$.
\item New Physics (NP): a signal component with $\vec \mu=(1.5,1.5)$ and covariance matrix $\Sigma = 0.1\,\mathbf{1}_{2\times 2}$ is present in addition to the background (reference) one. The expected signal is $N({\rm{S}})=500$ and the total number of expected events is $N({\rm NP})=N({\rm{S}})+N({\rm{R}})=10500$, with the remaining $10^4$ events generated by the reference model.
\end{itemize}
For a  2-5-1 network, Weight Clipping equal to $1.35$ and 150000 epochs, the results of our method are displayed in Figure~\ref{fig:DS} (bottom). We generated 1000 toy SM samples and 300 NP samples. The median significance, obtained with a $\chi^2$ approximation of the test statistic, is $20\sigma$, while Ref.~\cite{Hajer:2018kqm} quotes between $5$ and $16\sigma$ for the nearest neighbors approach depending on the cut on their discriminating variable. We can conclude that both approaches are sensitive to the simple problem at hand.

%%%%%%%%%%%%%%%%%%%%%%%%%%%%%%%%%%%%%%
\begin{figure}[!t]
\begin{center}
\includegraphics[width=0.9\textwidth]{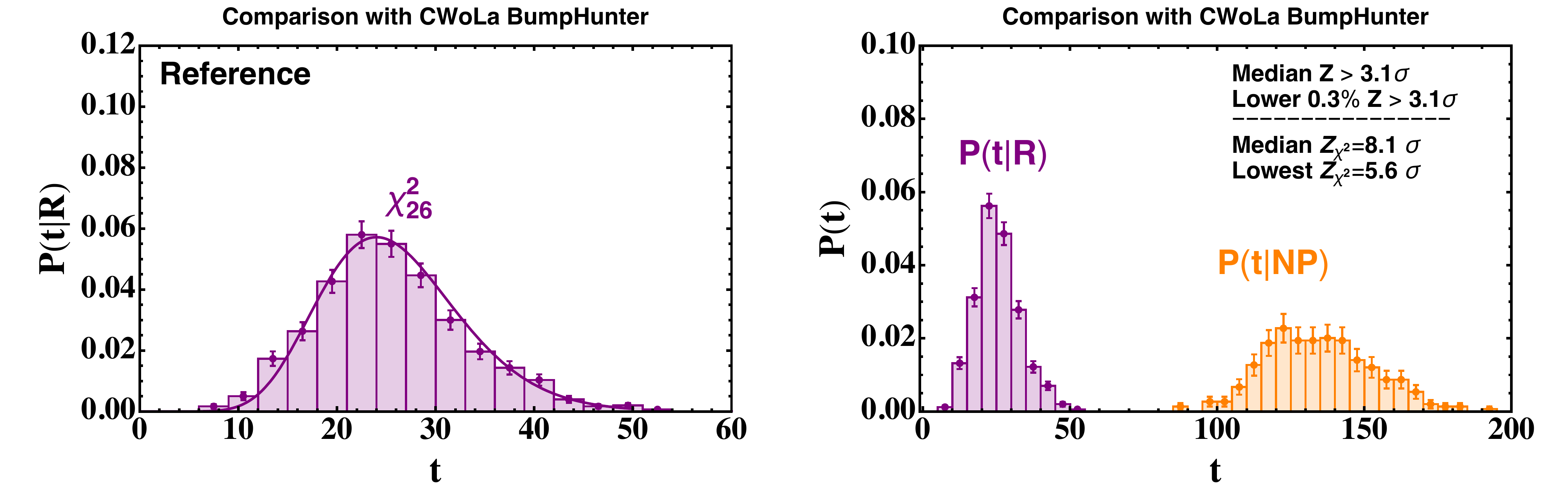}
\caption{Test statistic distribution for the signal and background models considered in~\cite{Collins:2018epr, Collins:2019jip}, obtained with our analysis technique. The figure refers to the 3D toy example discussed in v1 and v2 of~\cite{Collins:2018epr, Collins:2019jip}.}\label{fig:cwola}
\end{center}
\end{figure}
%%%%%%%%%%%%%%%%%%%%%%%%%%%%%%%%%%%%%%

The other idea that we compare with is the bump hunter technique in Ref.~\cite{Collins:2018epr, Collins:2019jip}. This approach does not have the same goal as ours, as it requires prior knowledge of the signal showing up as a peak in a pre-specified variable. It is further assumed that the background distribution of the other variables used in the analysis is the same in the peak and in the sideband regions. Clearly we have a price to pay in sensitivity for signals that satisfy these assumptions, since we discard this knowledge. On the other hand the approach in~\cite{Collins:2018epr, Collins:2019jip} is much less effective on (or blind to) signals that do not satisfy them. Given these differences, it is instructive to check what is exactly the price that we are paying on resonant signals compared to this refined bump hunter. 

We test our strategy on a three-dimensional toy example (see~\cite{Collins:2018epr, Collins:2019jip}) defined by:
\begin{itemize}
\item Reference model: the three variables $m, x$ and $y$ are uniformly distributed in the ranges $|m|<2$, $|x|<0.5$ and $|y|<0.5$. The number of expected events is $N({\rm R})=10000$.
\item New Physics (NP): there are $N({\rm S})=300$ signal events with variables uniformly distributed in the ranges: $|m|<1$, $|x|<0.1$ and $|y|<0.1$ and $N({\rm R})=10000$ events distributed as the reference model, for a total number of expected events $N({\rm NP})=N({\rm S})+N({\rm R})=10300$.
\end{itemize}
%The ideal significance quoted in~\cite{Collins:2018epr} is approximately 15$\sigma$. 
Our results are shown in Figure~\ref{fig:cwola} for a 3-5-1 network with Weight Clipping $3.4$ and 150000 epochs of training. As in the previous examples, we generated 1000 toy SM samples and 300 NP samples. 

We obtain a median significance of $8.1\sigma$, to be compared with the $10.8\sigma$ claimed in Ref.~\cite{Collins:2018epr, Collins:2019jip} for the optimal choice of the neural network discriminant threshold. In the comparison it should be taken into account that $10.8\sigma$ is a local significance, based on prior knowledge of the peak position and width. The degradation due to the need of scanning over the peak position and width (inherent of the bump hunter approach) and over the neural network threshold (specific of this strategy, see Ref.~\cite{Collins:2019jip}) should be quantified for a better comparison with our $8.1\sigma$ significance, which is instead global. However such a refined comparison is unnecessary in this benchmark example because no quantitative meaning should be attached to the asymptotic estimates of such high levels of significance. We can only conclude that our method is sensitive to this toy problem in spite of not being optimized for (and hence limited to) the detection of resonant signals.

%%%%%%%%%%%%%%%%%%%%%%%%%%%%%%%
\section{Benchmark Examples}\label{sec:datasets}
In the previous sections we have introduced our methodology and compared our data analysis strategy with two alternative ideas present in the literature. The comparisons involved toy examples that can not be directly mapped on cases of physical interest. 

The natural next step is to study the performances of our strategy on more realistic datasets and new physics examples. We choose to study LHC di-muon production and to consider two well-known new physics scenarios. In this section we describe the signal and background samples used for the analysis. The results of our study are presented in the next section. We consider two distinct possibilities for how new physics can manifest itself a resonant signal, represented by a $Z^\prime$ decaying to $\mu^+\mu^-$, and a smooth signal given by a contact interaction that we call ``EFT''. The samples used to study our performances are:

%%%%%%%%%%%%%%%%%%%%%%%%%%%%%%%%%%%%%%%%%%%%%%%%%%%%%%%%
%%%%%%%%%%%%%%%%%%%%%%%%%%%%%%%%%%%%%%%%%%%%%%%%%%%%%%%%
\begin{figure}[!t]
	\centering
	\includegraphics[width=0.8\linewidth]{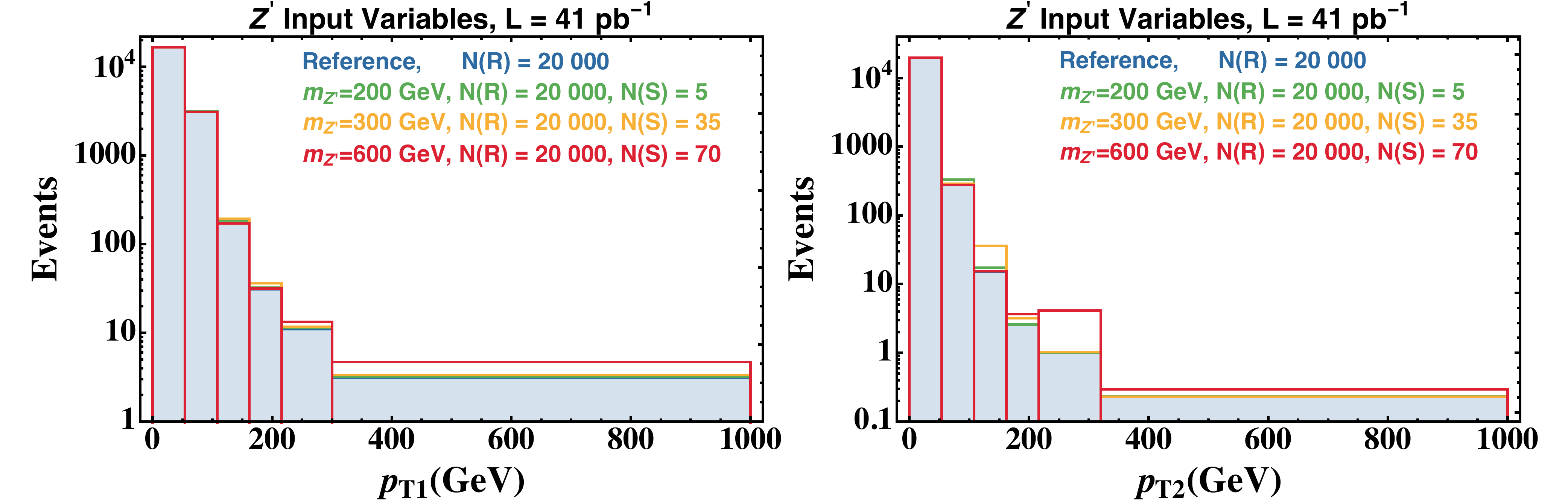} \\ \mbox{} \\
	\includegraphics[width=0.8\linewidth]{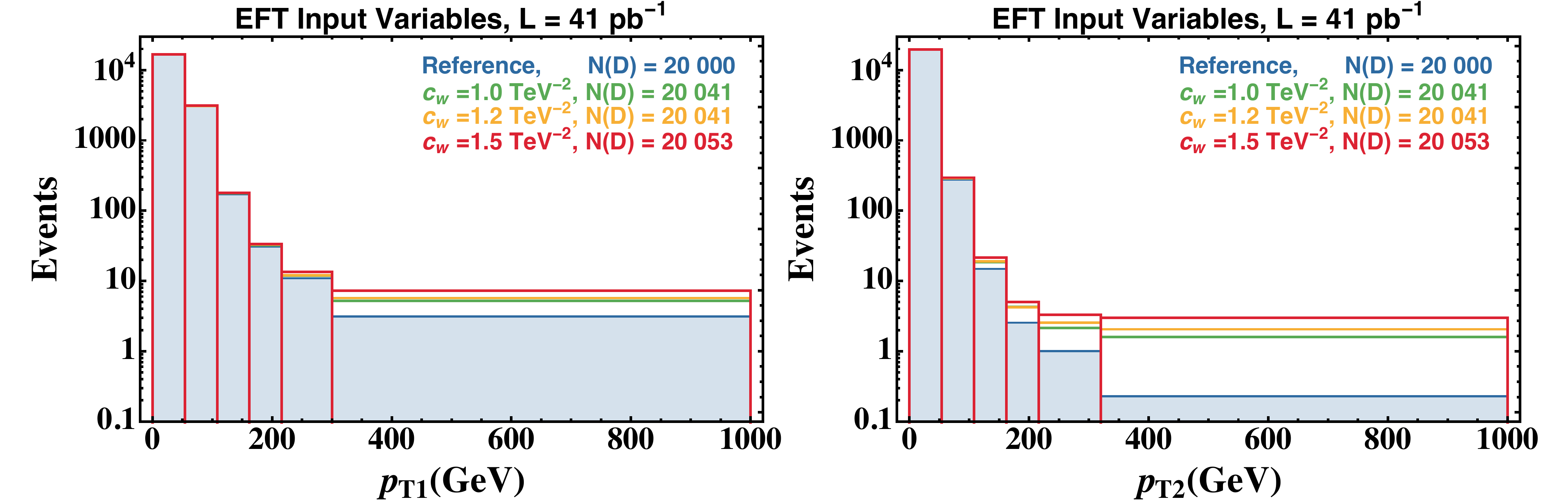}
	\caption{Transverse momenta of the two leading muons for a SM DY sample and our mock data samples containing a $Z^\prime$ decaying to muons (upper panel) or new physics events from the contact interaction in Eq.~\ref{eq:EFT} (lower panel). The samples are described in Section~\ref{sec:datasets}.}
	\label{fig:pT}
\end{figure}
\begin{figure}[!t]
	\centering
	\includegraphics[width=0.99\linewidth]{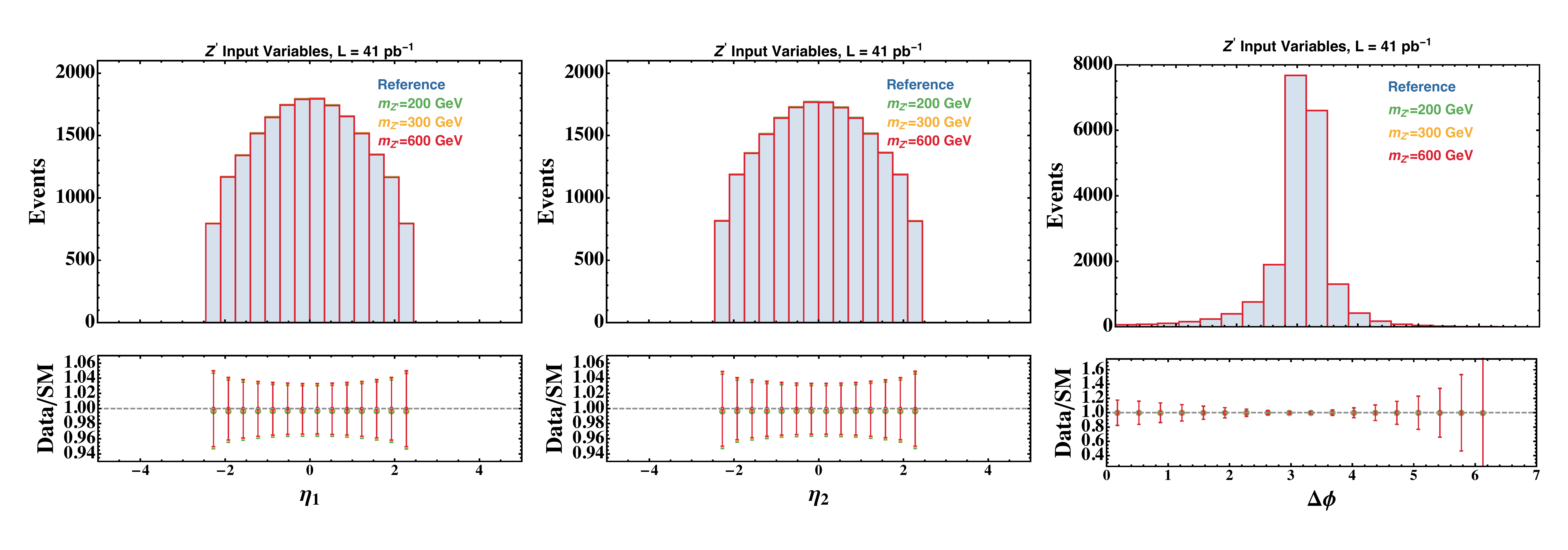} \\ \mbox{} \\
	\includegraphics[width=0.99\linewidth]{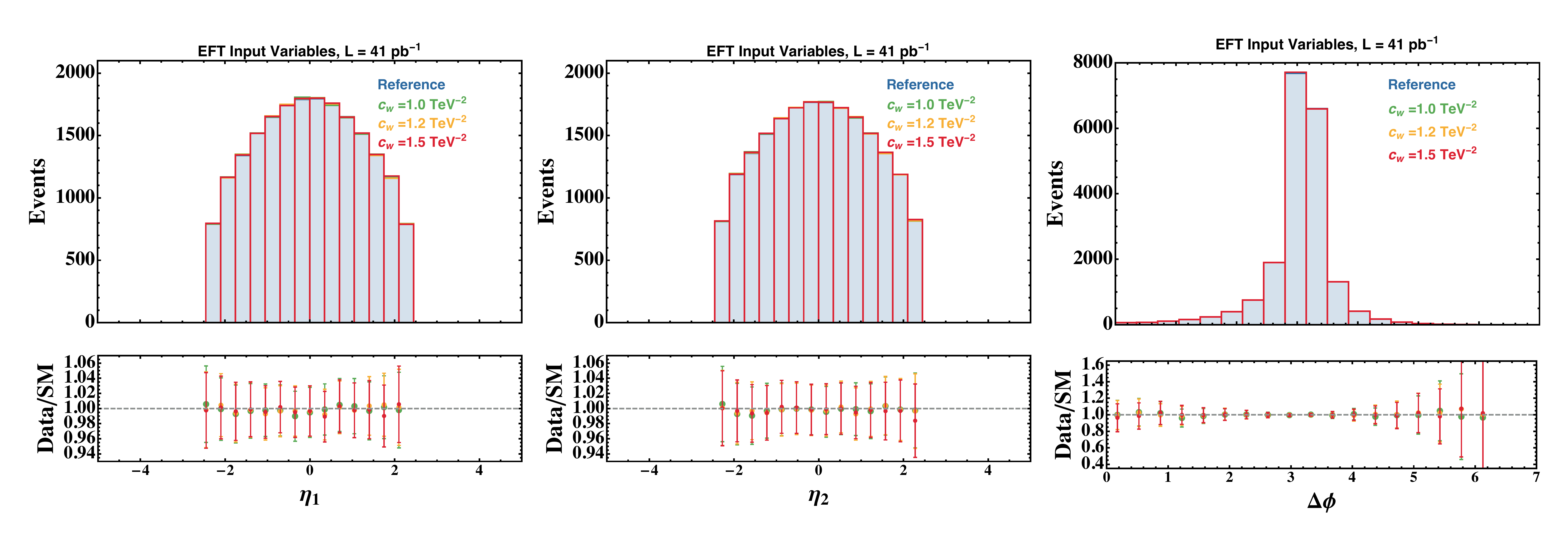}
	\caption{Pseudo-rapidities and $\Delta \phi$ of the two leading muons for a SM Drell-Yan sample and our mock data sample containing a $Z^\prime$ decaying to muons (upper panel) or new physics events from the contact interaction in Eq.~\ref{eq:EFT} (lower panel). The samples are described in Section~\ref{sec:datasets}.}
	\label{fig:angles}
\end{figure}
%%%%%%%%%%%%%%%%%%%%%%%%%%%%%%%%%%%%%%%%%%%%%%%%%%%%%%%%
%%%%%%%%%%%%%%%%%%%%%%%%%%%%%%%%%%%%%%%%%%%%%%%%%%%%%%%%

\paragraph{SM di-muon} The reference sample and the SM toy data are composed of SM Drell-Yan events: $pp\to \mu^+\mu^-$. All events were generated with {\tt MadGraph5}~\cite{Alwall:2011uj}, showered with {\tt Pythia6}~\cite{Sjostrand:2006za}, simulating proton-proton collision at $\sqrt{s}=13$~TeV with an average of 20 overlapping collisions per bunch crossing. The events were further processed with {\tt Delphes} 3~\cite{deFavereau:2013fsa}. We use the default CMS detector card. We run the {\tt Delphes} particle-flow algorithm, which combines the information from different detector components to derive a list of reconstructed particles. The five kinematical variables relevant for the analysis are the $p_T$'s and $\eta$'s of the two leptons and their $\Delta \phi$. These are given as input to the neural network after preprocessing. The integrated luminosity of the dataset, corresponding to the number of expected events in the toy SM (and BSM) samples are varied to study the performances of the algorithm as discussed in Section~\ref{sec:results}.

\paragraph{$\mathbf{Z^\prime}$ to di-muon} We study a new vector boson with the same couplings to SM fermions as the SM $Z$ boson. We generate events for three different masses: $m_{Z^\prime}=200, 300$ and $600$~GeV.  The signal manifest itself as a narrow resonance at the LHC: $\Gamma_{Z^\prime} \simeq \Gamma_Z m_{Z^\prime}/{m_Z}$. The events are generated using the same software and detector cards as the reference model events described above. The number of events in the data sample and the signal to background ratio $N({\rm S})/N({\rm R})$ are varied to study the performances of the algorithm and are discussed in Section~\ref{sec:results}. The input variables distribution for three representative signal points are shown in Figure~\ref{fig:pT} and Figure~\ref{fig:angles}.

\paragraph{EFT} We consider a non-resonant BSM effect due to the presence of a dimension-$6$ $4$-fermion contact interaction (see e.g. \cite{Farina:2016rws})
\be
%\mathcal{L}\supset - \frac{c_W}{4 m_W^2}\left(D_\rho W_{\mu\nu}^a\right)^2\; . 
\frac{c_W}{\Lambda^2}{J_L}_\mu^a {J_L}^\mu_a\,,
\label{eq:EFT}
\ee
where ${J_L}^\mu_a$ is the SU$(2)_L$ SM current and $\Lambda$ is conventionally set to $1$~TeV. We generate di-muon events with the same tools described above (supplemented with a {\tt MadGraph5} model for the EFT operator obtained by {\tt FeyRules}~\cite{Alloul:2013bka}) by varying $c_W$ in order to study the performances of the algorithm as discussed in Section~\ref{sec:results}. The distribution of the input values for three representative values of $c_W$ are shown in Figure~\ref{fig:pT} and Figure~\ref{fig:angles}.
\\
\mbox{}
\\
%The examples described above cover three of the most common new physics scenarios in QFT: a resonant signal in the bulk of a kinematical distribution ($m_{Z^\prime}=200, 300$~GeV), a resonant signal in its tail ($m_{Z^\prime}=600$~GeV) and a broad excess in the tail (EFT). By studying them we can show that our method is not sensitive to the details of the new physics present in the data. The results of ours study are presented in Section~\ref{sec:results}. In the next subsection we introduce a measure of the signal significance needed to quantify the performances of a model-independent search.
In Section~\ref{sec:results} we will extensively study the sensitivity of our method to the BSM scenarios described above. For a meaningful presentation of the results, and in order to compare the performances on different scenarios, we need an absolute measure of how much a given BSM hypothesis is ``easy'' to detect with a given integrated luminosity. As in Ref.~\cite{DAgnolo:2018cun}, this measure is introduced by the notion of ``ideal significance'', described below.

\subsection{The ideal significance}
The ideal significance is the highest possible median $Z$-score ($Z_{\rm id}$) that any search specifically targeted to a given BSM scenario in a given experiment could ever obtain. By the Neyman--Pearson lemma, it is obtained using the ``ideal'' test statistic 
\be
t_{\rm id}({\mathcal{D}})=2\,\log\left[\frac{e^{-N({\rm NP})}}{e^{-N({\rm{R}})}}\prod\limits_{x\in {\mathcal{D}}}\frac{n(x|{\rm NP})}{n(x|{\rm R})}\right]\,, \label{eq:teststatideal}
\ee
The ideal significance can be reached only in a fully model-dependent search where the exact knowledge of both the new physics distribution $n(x|{\rm NP})$ (see Table~\ref{tab:notation}) and the reference distribution $n(x|{\rm R})$ are available. This knowledge is available, in principle, for the BSM scenarios under examination. However computing $n(x|{\rm NP})/n(x|{\rm R})$ is cumbersome. An estimate for it sufficiently accurate to serve as a reference of performance in the present paper (hence quoted as $Z_{\rm ref}$) is readily obtained as follows.

\paragraph{$\mathbf{Z^\prime}$ to di-muon}
The signal shows up as a resonant peak in the di-muon invariant mass $m_{ll}$ around the $Z^\prime$ mass $m_{Z^\prime}$. A simple cut-and-count strategy in a suitably designed interval $m_{ll}\in\left[m_{\rm min}, m_{\rm max}\right]$ around $m_{Z^\prime}$ should provide a reasonable estimate of the ideal reach. The ideal significance is thus estimated as
\be
Z_{\rm ref} = Z\big[1-{\rm{CDF}[{\rm{P}}_b]}(s+b)\big] \sim \frac{s}{\sqrt{b}}\,,\;\;\;\;{\rm{where}}\;\;\left\{
\begin{array}{l}
s=f_{\rm sig} N({\rm S})\\
b=f_{\rm bkg}N({\rm R})
\end{array}\right. \,. \label{eq:idealZ}
\ee
In the equation, ${\rm{CDF}[{\rm{P}}_b]}$ denotes the cumulative of the Poisson distribution with mean ``$b$'' while $f_{\rm sig}$ and $f_{\rm bkg}$ are respectively the signal and background fractions in the mass-window
\be
f_{\rm sig} \equiv \int_{m_{\rm min}}^{m_{\rm max}} \hspace{-10pt} dm_{ll} \frac{dP(m_{ll} |{\rm S})}{dm_{ll}} \,,\;\;\;\;\;
f_{\rm bkg} \equiv\int_{m_{\rm min}}^{m_{\rm max}} \hspace{-10pt} dm_{ll} \frac{dP(m_{ll}|{\rm R})}{dm_{ll}}\,.\nonumber
\ee
The signal fraction is estimated with a Monte Carlo sample consisting of 16000 signal-only events. The background is computed by fitting a Landau distribution to the tail of a SM sample with 1.6 million events. The boundaries of the mass-window $\left[m_{\rm min}, m_{\rm max}\right]$ are selected by optimizing the significance and reported in Table~\ref{tab:Zideal} together with the corresponding signal and background fractions. 

%Notice that the the optimal window depends not only on the $Z^\prime$ mass, but also on the total signal-over-background ratio $N({\rm{S}})/N({\rm{R}})$. The benchmark values of $N({\rm{S}})/N({\rm{R}})$ considered in the table correspond to the most common configurations we will encounter in Section~\ref{sec:results}. Mild variations of $N({\rm{S}})/N({\rm{R}})$ relative to the benchmark, which are also considered in Section~\ref{sec:results}, do not change the optimal window appreciably.

\begin{table}[!t]
\begin{center}
\begin{tabular}{|c|c|c|c|c|c|}
$m_{Z^\prime}$ & $N({\rm{S}})$ & $N({\rm{R}})$ & $\left[m_{\rm min}, m_{\rm max}\right]$& $f_{\rm sig}$&$f_{\rm bkg}$\\
\hline
$200$ GeV & 40, 60, 80&	$20\cdot10^{3}$ & $\left[185,\,215\right]$&0.62&$9.0\times10^{-4}$\\
\hline
$300$ GeV & 20, 30&	$20\cdot10^{3}$ & $\left[278,\,322\right]$&0.72&$2.3\times10^{-4}$\\
& 25, 35&	$20\cdot10^{3}$ & $\left[279,\,321\right]$&0.71&$2.2\times10^{-4}$\\
\hline
$600$ GeV & 6, 10&	$20\cdot10^{3}$ & $\left[554,\,656\right]$&0.77&$2.4\times10^{-5}$\\
& 15&	$20\cdot10^{3}$ & $\left[549,\,662\right]$&0.83&$2.8\times10^{-5}$\\
\end{tabular}
\caption{Mass-window, signal and background fractions for the estimate of $Z_{\rm id}$ in eq.~(\ref{eq:idealZ}).}\label{tab:Zideal}
\end{center}
\end{table}

In order to validate eq.~(\ref{eq:idealZ}) as a reasonable estimate of $Z_{\rm id}$ we compared it with the truly ``ideal'' significance obtained with the Neyman-Pearson test performed on the $m_{ll}$ variable. We fitted the background Monte Carlo data using  two Landau distributions (one for $250\;{\rm GeV}\leq m_{ll}<600\; {\rm GeV}$, the other for $m_{ll}\geq 600\;{\rm GeV}$) and a Normal distribution for the $Z^\prime$ peak. This allowed us to compute the test statistic in eq.~(\ref{eq:teststatideal}) and in turn the ideal significance by toy experiments. Good agreement with eq.~(\ref{eq:idealZ}) was found. Notice however that the comparison was possible only in configurations with low enough $Z_{\rm ref}$. For cases with $Z_{\rm ref}\gtrsim4$, which we do consider in Section~\ref{sec:results}, validation is unfeasible and we exclusively rely on eq.~(\ref{eq:idealZ}).

\paragraph{EFT} Also in this case, the di-muon invariant mass is the most relevant discriminant. Since the excess is spread over the entire spectrum, the ideal significance is estimated through a Likelihood Ratio (Neyman--Pearson) test on the binned $m_{ll}$ distribution. The number of expected events in each bin is quadratic in $c_W$
\be\label{nicw}
n_i(c_W) = N({\rm{R}})(\alpha_i+\beta_i c_W+\gamma_i c_W^2) \,,
\ee
with coefficients determined from Monte Carlo simulations at varying $c_W$, reported in Table~\ref{tab:eft_fit}. The test statistic is the log-ratio for the Poisson distributed observed countings ``$o_i$'' in each bin
\begin{equation}
t({\mathcal{D}})=\sum\limits_{i\in{\mathrm{bin}}}2\left[n_i(0)-n_i(c_W)+o_i\log\frac{n_i(c_W)}{n_i(0)}\right]\,,
\end{equation}
and the reference significance is evaluated from the distribution of $t$ in the SM ($c_W=0$) extracted from toy experiments. 

%%%%%%%%%%%%%%%%%%%%%%%%%%%%%%%%%%%%%%%%%%%%%%%%%%%%%%%
\begin{table}[t]
\begin{center}
		\scalebox{0.65}{
		\centering
		\begin{tabular}{|l|ccccc|}
			\hline 
			%&&&$\sigma \,[{\rm pb}^{-1}]$&&\\
			& \bf{Bin 1}&  \bf{Bin 2}&\bf{Bin 3}&\bf{Bin 4}&\bf{Bin 5}\\ 
			& \bf{[60,148]GeV} & \bf{[148, 296] GeV}&\bf{[296, 444] GeV}&\bf{[444, 592] GeV}&\bf{[592, 740] GeV}\\ 
			\hline 
			%$-10^{-4}$&	520.108740&91.5761657&102.897670&93.6514880&83.7783742\\
			%\hline 
			%$-5\times10^{-5}$&	498.190828&18.3546262&23.7597739&23.3746168&20.4694149\\ 
			%\hline 
			%$0$&		469.711702&4.35498510&0.231245085&0.0466900978&0.0140335785\\ 
			%\hline 
			%$10^{-6}$&	496.031871&2.73581526&0.284793759&0.0849209073&0.0385223919\\ 
			%\hline 
			%$5\times10^{-5}$&	509.080522&6.19460807&2.06534365&1.38331496&2.46933704\\ 
			%$10^{-5}$&	526.163684&43.4287392&31.4612374&25.8407284&22.1597597\\ 
			%$10^{-4}$&	593.732382&141.225028&116.876227&100.709347&85.9897370\\ 
			%\hline
			%$\alpha$	&$496.7  \pm 2.8 \,(0.55\%)$&	$2.48  \pm0.02  \, (0.75\%)$&	$0.222  \pm0.008 \,(3.46\%)$&	$0.0467  \pm 0.0005\,  (1.1\%)$&	$(0.01403 \pm 0.00008 ) \, (0.54\%)$\\
			%$\beta\,[{\rm TeV^{2}]$&$(34.8  \pm 2.7)10^{5} \, (7.8\%)$&	$(252.2 \pm 1.9)10^3 \,(0.76\%)$&	$(69.1\pm 6.4)10^3  \,(9.3\%)$&	$(28.3 \pm1.1)10^3   \,(3.7\%)$& $(159.8  \pm 2.2)10^2\, (0.41\%)$\\
			%$\gamma\,[{\rm TeV}^{4}]$& $(60.1 \pm 4.5)10^8 \,(7.4\%)$&	$(1139.8 \pm 3.5)10^7  \,(0.30\%)$&	$(109.7\pm 2.0)10^8 \,(1.9\%)$& $(977.1 \pm 5.5)10^7 \, (0.6\%)$& $(850.4 \pm 2.3)10^7 \,    (0.27\%)$\\
			
			$\alpha$&		$(9.93 \pm 0.03)10^{-1} \,(0.3\%)$	&$(4.95 \pm 0.03)10^{-3} \,(0.6\%)$	&$(4.01 \pm 0.02 )10^{-4}\,(0.5\%)$	&$(8.5 \pm 0.2)10^{-5}\,  (2.3\%)$	&$(2.64 \pm 0.02 )10^{-5} \, (0.8\%)$\\
			$\beta$&		$(7.0 \pm 0.5)10^{-4} \, (3.5\%)$	&$(5.06 \pm 0.04)10^{-4} \,(0.8\%)$	&$(1.52 \pm 0.02)10^{-4}  \,(1.3\%)$	&$(6.7 \pm 0.3)10^{-5}   \,(4.4\%)$	&$(3.37  \pm 0.05)10^{-5}\, (1.5\%)$\\
			$\gamma$& 	$(1.21 \pm 0.07)10^{-5} \,(5.8\%)$	&$(2.28 \pm 0.07)10^{-5}  \,(3.0\%)$	&$(2.20\pm 0.07)10^{-5} \,(3.1\%)$	&$(1.96 \pm 0.02)10^{-5} \,(1.0\%)$	&$(1.703 \pm 0.005)10^{-5} \,    (0.29\%)$\\
			\hline \hline
			%&&&$\sigma \,[pb^{-1}]$&&\\
			& \bf{Bin 6}&  \bf{Bin 7}&\bf{Bin 8}&\bf{Bin 9}&\bf{Bin 10}\\ 
			&\bf{[740, 888] GeV}&\bf{[888, 1241] GeV}&\bf{[1241, 1594] GeV}&\bf{[1594, 1947] GeV}&\bf{[1947, 2300] GeV}\\ 
			\hline 
			$\alpha$&					$(9.84 \pm 0.05)10^{-6} \,(0.5\%)$	&$(7.59 \pm 0.07)10^{-6} \,(0.9\%)$	&$(1.4 \pm 0.2 )10^{-6}\,(14\%)$	&$(3.6 \pm 0.7)10^{-7}\,  (19\%)$	&$(1.10 \pm 0.03 )10^{-7} \, (0.8\%)$\\
			$\beta$&		$(1.85 \pm 0.01)10^{-5} \, (0.5\%)$	&$(2.30 \pm 0.01)10^{-5} \,(0.4\%)$	&$(6 \pm 1)10^{-6}  \,(17\%)$	&$(2.2 \pm 0.6)10^{-6}   \,(27\%)$	&$(1.34  \pm 0.09)10^{-6}\, (1.5\%)$\\
			$\gamma$& 	$(1.432 \pm 0.005)10^{-5} \,(0.3\%)$	&$(2.90 \pm 0.01)10^{-5}  \,(0.3\%)$	&$(1.6 \pm 0.2)10^{-5} \,(13\%)$	&$(1.09 \pm 0.05)10^{-5} \,(4.6\%)$	&$(6.769 \pm 0.001)10^{-6} \,    (0.01\%)$\\
			\hline
		\end{tabular}
			}
			
		%\scalebox{0.65}{
		%\centering		
		%\begin{tabular}{|l|ccccc|}
						%$-10^{-4}$&	70.5186815&144.462686&95.6541346&59.2532015&34.4916208\\
			%\hline 
			%$-5\times10^{-5}$&	17.4643696&35.8485705&23.4807692&14.3057189&8.45506247\\ 
			%\hline 
			%$0$&		0.00515804791&0.00426119235&0.000818722902&0.000201663796&0.0000566030448\\ 
			%\hline 
			%$10^{-6}$&	0.0210833541&0.0298792899&0.00791858627&0.00399429517&0.00204297966\\ 
			%\hline 
			%$5\times10^{-5}$&	0.817262197&1.53464727&0.986583546&0.595492151&0.343150624\\ 
			%$10^{-5}$&	18.4410362&37.0245694&23.8440689&14.7747093&8.17661712\\ 
			%$10^{-4}$&	72.0944538&145.556441&94.8327319&58.9134974&31.2966615\\ 
			%\hline
			%$\alpha$	& $(515.5 \pm3.3)10^{-5}\, (0.63\%)$ & $(426.3\pm4.6)10^{-5}\, (1.1\%)$	&$(81.8  \pm2.4)10^{-5}   \, (2.9\%)$	&$(201.6\pm5.9)10^{-6}\, (2.9\%)$	&$(56.6\pm2.3)10^{-6} \,(4.1\%)$\\
			%$\beta\,[{\rm TeV}^{2}]$	& $8816.9 \pm138.3 \,(1.6\%)$ & $(11020  \pm 338.3) \, (3.1\%)$		&$-2471.6 \pm 240.6 \, (9.7\%)$	&$-2110  \pm130.6 \,  (6.2\%)$	&$-1409.1  \pm117.7 \,  (8.4\%)$	\\
			%$\gamma\,[{\rm TeV}^{4}]$& $(716.7 \pm 2.114)10^7  \,(0.30\%)$ & $(1448.2  \pm 7.1)10^7 \, (0.49\%)$	&$(96.0  \pm1.3)10^8  \,(1.3\%)$	&$(591.3\pm7.8)10^7 \,(1.3\%)$	&$(340.0      \pm8.6)10^7\, (2.5\%)$	\\

		%	\hline
		%\end{tabular}
	%}
	\caption{The coefficients of the polynomial fit in eq.~(\ref{nicw}).}
\label{tab:eft_fit}
\end{center}
\end{table}

%%%%%%%%%%%%%%%%%%%%%%%%%%%%%%%
\section{Results on Benchmark Examples}\label{sec:results}
\begin{figure}[!t]
		\centering
		\includegraphics[width=0.99\linewidth]{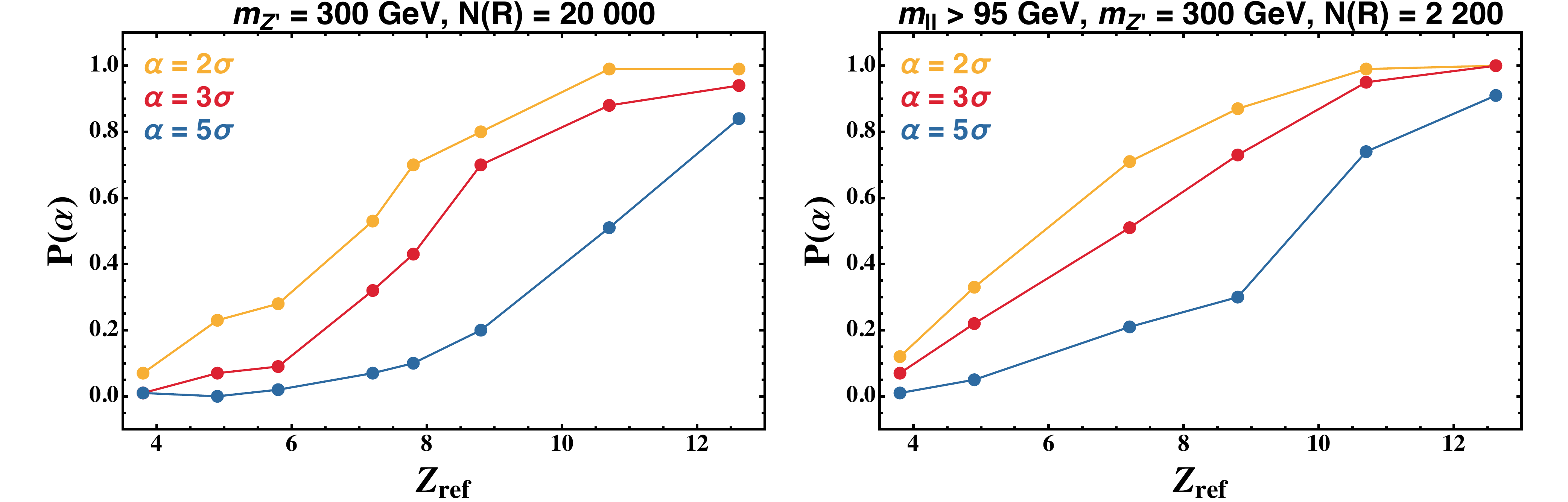}
	\caption{Probability of finding a $\alpha=2\sigma, 3\sigma, 5\sigma$ evidence for new physics using our technique as a function of the reference significance of the signal, for the $Z^\prime$ model described in Section~\ref{sec:datasets}. (Left) Including the $Z$-peak in the data. (Right) Without the $Z$-peak.}
	\label{fig:sensitivity}
\end{figure}

\begin{figure}[!t]
		\centering
		\includegraphics[width=0.99\linewidth]{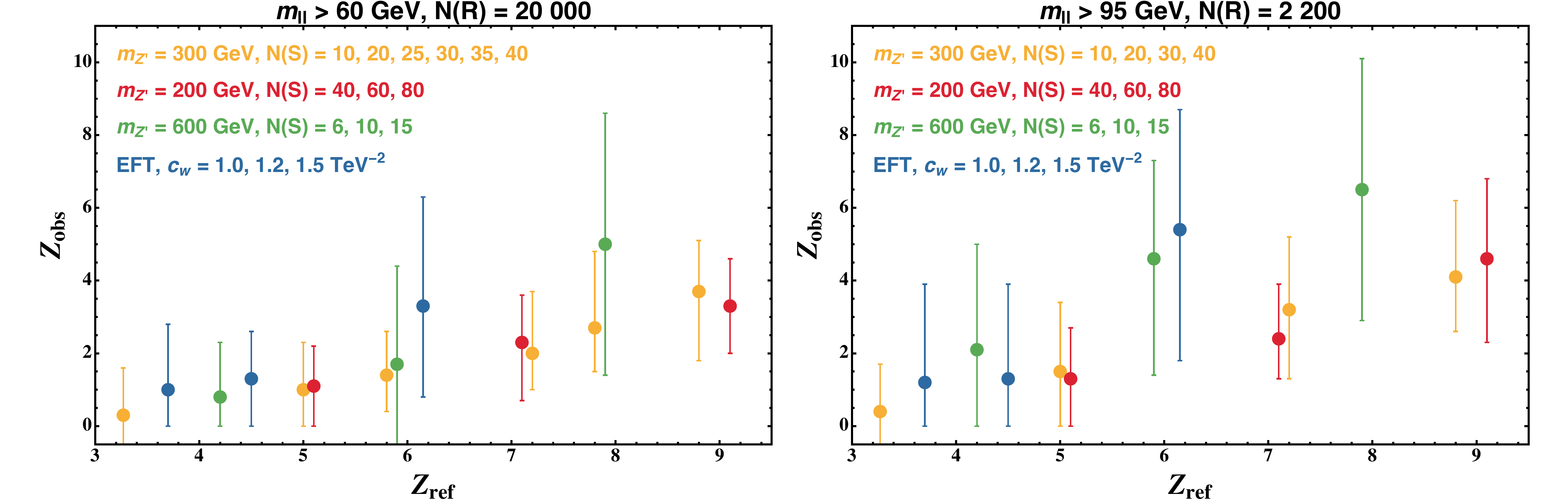}\\
	\caption{Sensitivity ($Z_{\rm obs}$) to $Z^\prime \to \mu^+\mu^-$ for $m_{Z^\prime}=300$~GeV and the EFT signal. We show the sensitivity as a function of the reference significance $Z_{\rm ref}$.}
	\label{fig:mInd}
\end{figure}

\begin{figure}[!t]
		\centering
	$ \begin{array}{cc}
		\includegraphics[width=0.9\linewidth]{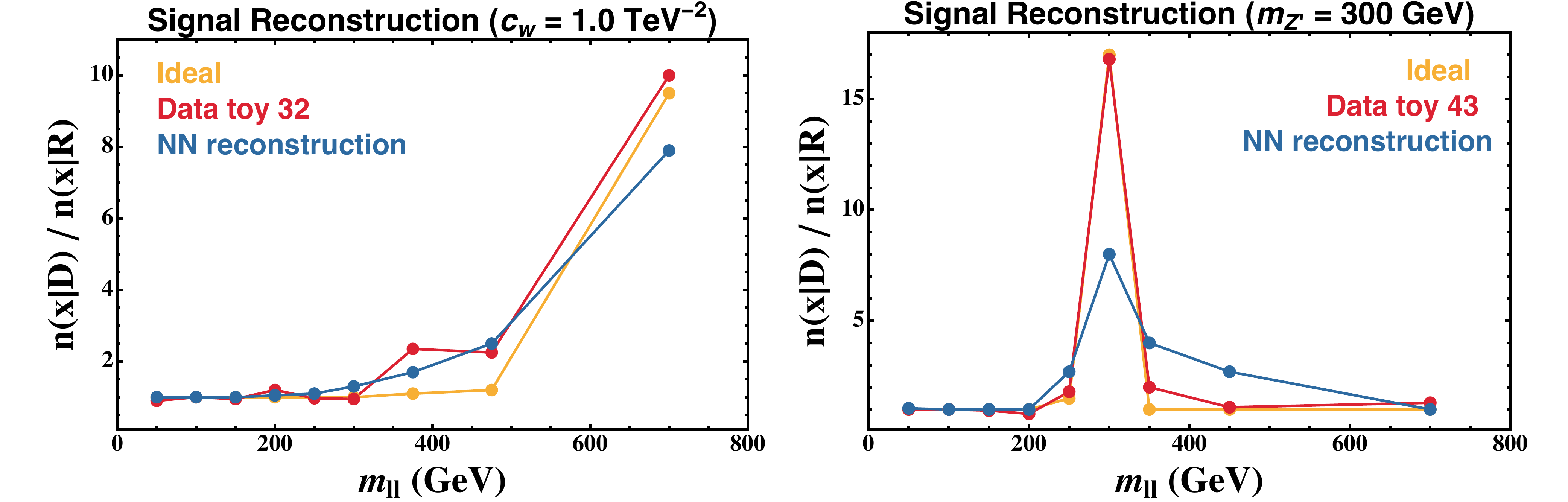} &
	\end{array} $
	\caption{Comparison between the ideal invariant mass distribution for the $Z^\prime$ and EFT signals and the distribution reconstructed by the network and realized in the toy sample taken as input. The probability distribution of the data sample is normalized to the reference one.}
	\label{fig:reco}
\end{figure}

\begin{figure}[!t]
		\centering
		\includegraphics[width=0.9\linewidth]{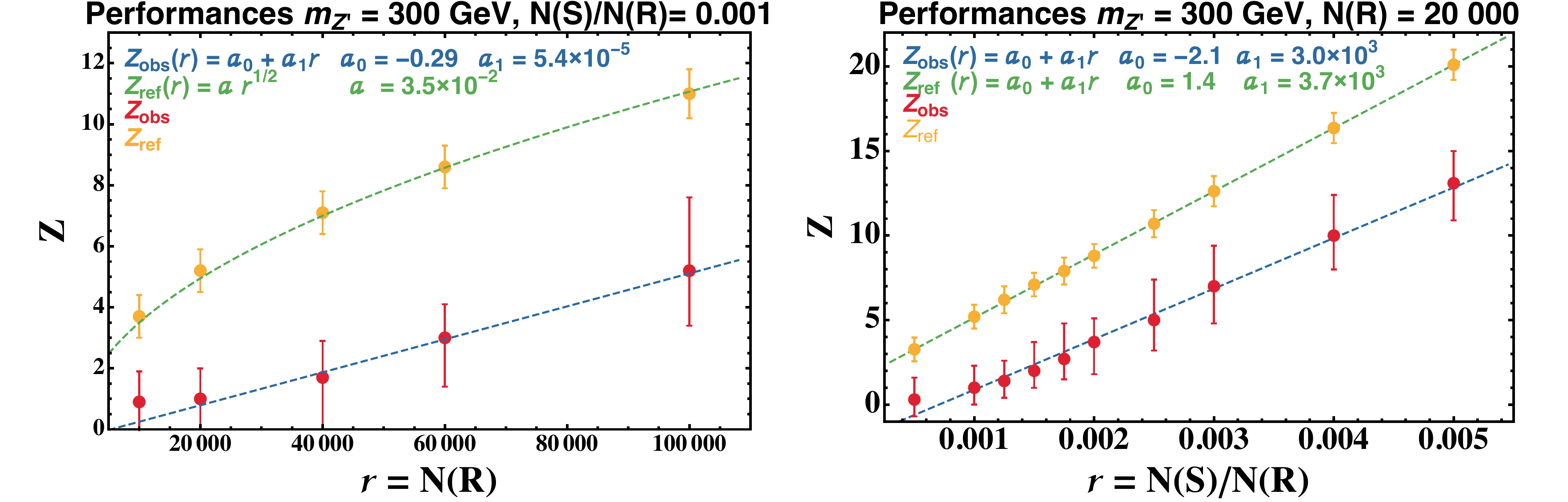}
	\caption{Sensitivity ($Z_{\rm obs}$) to $Z^\prime \to \mu^+\mu^-$ for $m_{Z^\prime}=300$~GeV. We show the sensitivity as a function of Luminosity (left panel) and signal fraction (right panel). For reference we plot the reference significance $Z_{\rm ref}$ and a polynomial fit to the sensitivity.}
	\label{fig:lumiscale}
\end{figure}

In this section we study the sensitivity of our data analysis strategy to the physics examples discussed in the previous section. Our main results are:
\begin{enumerate}
\item In the examples we studied, in all cases where the estimate of the ideal significance exceeds $5\sigma$, the probability of finding a $2\sigma$ tension for the SM using our approach is $p(\alpha=2\sigma)\gtrsim 20\%$ and grows to $p(\alpha=2\sigma)\gtrsim 40\%$ if we exclude the $Z$-boson peak from the input data by a cut on the invariant mass. The probability of finding a $3\sigma$ tension is $p(\alpha=3\sigma)\gtrsim 7\%$ and $p(\alpha=3\sigma)\gtrsim 20\%$ including or excluding the $Z$-peak, respectively. [Figure~\ref{fig:sensitivity}]
\item For any given ``experiment'' (i.e., at fixed luminosity and input space), the observed significance mostly depends on the ideal significance of the putative signal, while it weakly depends on the type of signal. [Figure~\ref{fig:mInd}] 
\item The neural network output correctly reconstruct the data to reference likelihood-ratio, finding a good approximation to the properties of the signal in the space of input variables, for all the signals that we consider. [Figure~\ref{fig:reco}]
\item In the examples that we have studied, where the observed significance is not close to saturating the reference significance, the observed significance increases linearly with luminosity, as opposed to the $\sqrt{L}$ growth of the reference significance. Both significances increase linearly with the number of signal events as expected. [Figure~\ref{fig:lumiscale}]
%\item The observed significance increases linearly with luminosity, as opposed to the $\sqrt{L}$ growth of the ideal significance. It increases linearly with the number of signal events as expected. [Figure~\ref{fig:lumiscale}]
\end{enumerate}
Properties ``1'' and ``2'' make our technique ideally suited to identify an unexpected new physics signal. Because of ``3'', if a tension is observed in the data the sensitivity to the signal can be increased with a dedicated analysis on new data, selected using the likelihood ratio learned by the network.

As stated in ``2'' above, $Z_{\rm obs}$ essentially depends only on the ideal significance (approximated by $Z_{\rm ref}$) for a given experiment. However in a different experiment (e.g., if we change the luminosity) the relation between $Z_{\rm obs}$ and $Z_{\rm ref}$ changes. The relation becomes more favorable at high luminosity because of point ``4''.

Let us now turn to an extensive description of the items above, and of our findings on a few technical points relevant to the implementation of the algorithm. For all the results in this paper the minimization of the loss function is performed using ADAM~\cite{Kingma:2014vow} as implemented in {\tt Keras}~\cite{Keras} (with the {\tt TensorFlow}~\cite{TensorFlow} backend) with parameters fixed to: $\beta_1=0.9$, $\beta_2=0.99$, $\epsilon=10^{-7}$, initial learning rate $=10^{-3}$. The batch size is always fixed to cover the full training sample. Network architecture, size of the weight clipping and number of training rounds were selected following the procedure described in Section~\ref{sec:method}. Where not specified otherwise, the results were obtained with a 5-5-5-5-1 network and $3\times 10^5$ training rounds, using 100 data samples and 100 toy reference samples. The median observed significance plotted in the Figures and its $68\%$ C.L. error were obtained approximating $P(t|R)$ with a $\chi^2$ distribution with as many degrees of freedom as free parameters in the network as discussed in Section~\ref{sec:method}. We always consider a five-dimensional input space composed of the $p_T$'s and $\eta$'s of the two leptons and their $\Delta \phi$. The range of the input variables and their distribution for three representative signal points are shown in Figure~\ref{fig:pT} and Figure~\ref{fig:angles}. 

%%%%%%%%%%%%%%%%%%%%%%%%%%%%%%%%%%%%%%%%%%%%%%%%%%%
%%%%%%%%%%%%%%%%%%%%%%%%%%%%%%%%%%%%%%%%%%%%%%%%%%%
%%%%%%%%%%%%%%%%%%%%%%%%%%%%%%%%%%%%%%%%%%%%%%%%%%%
%%%%%%%%%%%%%%%%%%%%%%%%%%%%%%%%%%%%%%%%%%%%%%%%%%%

\paragraph{Sensitivity} The first goal of our study is to show that our technique is sensitive to realistic signals. By realistic we mean having $N({\rm{S}})/N({\rm{R}})\ll 1$, i.e. a small number of signal events compared to the total size of the sample, and ideal significances of order a few $\sigma$'s. These choices reproduce signals that we might have missed at the LHC so far, if not targeted by a dedicated search. The best way to illustrate the performances of a model-independent strategy is to report the probability it has to identify a tension with respect to the SM if a putative new physics effect is present in the data. This measures the chances that the analysis has to produce an interesting result. In the left panel of Figure~\ref{fig:sensitivity} we show the probability of finding evidence for new physics at the $\alpha=2\sigma, 3\sigma$ and $5\sigma$ levels given a reference significance for the signal. We consider for illustration the $Z^\prime$ signal model with $m_{Z^\prime}=300$~GeV described in the previous section, but similar or better performances are obtained for other masses and for the case of the EFT. The two plots here presented are obtained by fixing the luminosity while the ratio  $N({\rm{S}})/N({\rm{R}})$ is varied. %The signal fraction is fixed to $N({\rm{S}})/N({\rm{R}})=10^{-3}$, the size of the reference sample is $\mathcal{N}_R=5\,N({\rm{R}})$ and we increase $N(R)$ from $10^4$ to $10^5$. 

On the left panel of the figure, and in the results that follow if not specified otherwise, we applied our algorithm to the entire dataset which includes the SM $Z$-boson peak. This choice was made in order to challenge our analysis strategy in a situation where the dataset is dominated by the peak, where no new physics effect is present. On the other hand the peak would be excluded in a realistic application of our method to the di-muon final state because it is hard to imagine new physics appearing on the $Z$ peak not excluded by LEP and because detailed analyses of the $Z$ resonant production could be performed separately. If we exclude the $Z$-peak from the input data, with a cut $m_{ll}>95$~GeV (whose efficiency is $10\%$), the performances of our analysis improve as shown on the right panel of Figure~\ref{fig:sensitivity}.

Another way to quantify the sensitivity is to report the median significance obtained for different new physics scenarios, still as a function of the ideal significance. The result is shown in Figure~\ref{fig:mInd}, with the error bars representing the $68\%$~C.L. spread of the observed significance distribution. The study was performed for a given experimental setup, namely by fixing $N({\rm{R}})=2\times 10^4$ (and  $\mathcal{N}_R=5\,N({\rm{R}})$), and varying the signal fraction or the EFT Wilson coefficient $c_W$ as shown in the legend. We observe, similarly to Ref.~\cite{DAgnolo:2018cun}, a good level of correlation between our sensitivity and the ideal one and a weak dependence on the nature of the new physics. This correlation was sharper in the examples studied in Ref.~\cite{DAgnolo:2018cun}, however it should be taken into account that the present study relies on approximate (see Section~\ref{sec:datasets}) estimates of $Z_{\rm{id}}$ and that high values of $Z_{\rm{obs}}$ are also approximate, being estimated with the Asymptotic $\chi^2$ formula (see Section~\ref{sec:alg}). 

%Furthermore the strongest departure from a universal behavior $Z_{\rm{obs}}\simeq f(Z_{\rm{id}})$ is observed for the  $600$~GeV $Z^\prime$ signal, which emerges in a phase-space region which is extremely rare in the reference model. It is possible to argue that the significance cannot behave universally in these conditions. It is thus conceivable that the nearly universal behavior $Z_{\rm{obs}}\simeq f(Z_{\rm{id}})$ we observe might not be an accident, but rather have a deep explanation which we could not identify.

%%%%%%%%%%%%%%%%%%%%%%%%%%%%%%%%%%%%%%%%%%%%%%%%%%%
%%%%%%%%%%%%%%%%%%%%%%%%%%%%%%%%%%%%%%%%%%%%%%%%%%%
%%%%%%%%%%%%%%%%%%%%%%%%%%%%%%%%%%%%%%%%%%%%%%%%%%%
%%%%%%%%%%%%%%%%%%%%%%%%%%%%%%%%%%%%%%%%%%%%%%%%%%%

\paragraph{Likelihood Learning} It is instructive to study directly what the network has learned during training. The network should learn approximately the log-ratio between the true distribution ($n(x|{\rm{T}})$, see Table~\ref{tab:notation}) of the data and the reference model distribution $n(x|{\rm{R}})$. We should thus be able to get information on the nature of the discrepancy by inspecting the likelihood ratio learned by the network as a function of the physical observables chosen as input or any of their combinations. In the case of a $Z^\prime$ signal, for instance, we would like to see a bump in the invariant mass distribution as learned by the network. 

In Figure~\ref{fig:reco} we plot the distribution ratio learned by the network as a function of the invariant mass of the dimuon system. In the Figure we also show the true likelihood ratio used for the generation of the events and its estimate based on the specific data sample used for training. The signals are the $Z^\prime$ with a $300$~GeV mass with $N({\rm{S}})/N({\rm{R}})=2\times10^{-3}$, $N({\rm{R}})=2\times10^4$ and $\mathcal{N}_R=10^5$ and an EFT signal with the same $N({\rm{R}})$ and $\mathcal{N}_R$ and $c_W=10^{-6}$. Notice that $m_{ll}$ is not given to the network, the input variables being the muon $p_T$'s, rapidities and $\Delta\phi$.

%%%%%%%%%%%%%%%%%%%%%%%%%%%%%%%%%%%%%%%%%%%%%%%%%%
%%%%%%%%%%%%%%%%%%%%%%%%%%%%%%%%%%%%%%%%%%%%%%%%%%
%%%%%%%%%%%%%%%%%%%%%%%%%%%%%%%%%%%%%%%%%%%%%%%%%%
%%%%%%%%%%%%%%%%%%%%%%%%%%%%%%%%%%%%%%%%%%%%%%%%%%

The ratios in the figure were obtained in the following way. The yellow ``ideal" likelihood-ratio was obtained by binning the invariant mass of a large data sample, containing one million events, and of the reference sample and taking the ratio. The red likelihood-ratio pertaining to a specific toy was obtained in the same way, replacing the large data sample with the relevant toy. Finally, the ratio as learned by the network was obtained by reweighting reference sample by $e^{f(x, {\widehat {\bf w}})}$, where $f$ is the neural network output after training, binning it and taking the ratio with the reference.

The network is doing a pretty good job in reproducing a peak or a smooth growth (for the $Z^\prime$ and the EFT, respectively) in the invariant mass. Therefore if one had access to a new independent data set, distributed like the one used for training (i.e., following $n(x|{\rm{T}})$), one could employ the neural network $f(x, {\widehat {\bf w}})$ (trained on the first dataset) as discriminant (for instance, by a simple lower cut), and boost the significance of the observed tension.

In the studies presented so far we have chosen as input to the network five independent kinematic variables that characterize the di-muon final state under examination, paying attention not to include the invariant mass $m_{ll}$ which is essentially the only relevant discriminant in the new physics scenarios under investigation. This choice was intended to maximize the difficulty of the network task, reproducing the realistic situation where, since the actual signal is unknown, the most discriminant variable cannot be identified and given to the network. However it is interesting to study the potential improvement of the performances that could be achieved with a judicious (but model-dependent) choice of the input variables. The first test we made was to present $m_{ll}$ to the network in addition to the five variables $p_{T1,2}$, $\eta_{1,2}$ and $\Delta\phi$. This led to no substantial improvement of the performances suggesting that the neural network is already learning to reconstruct $m_{ll}$ sufficiently well from the five variables and does not need the sixth one. The second test was to trade the variable $\Delta\phi$ for $m_{ll}$, considering an alternative five-dimensional parametrization of the phase-space. Notice that $\Delta\phi$ has no discriminating power whatsoever because the new physics scenarios under examination emerge in $2\to2$ scattering processes where the muons are back-to-back in the transverse plane up to showering and detector effects, as it is the case for the SM. The $\Delta\phi$ distribution is thus (see Figure~\ref{fig:angles}) strongly peaked at $\pi$ and identical in the SM and in BSM. Replacing it with $m_{ll}$, which is instead the most discriminant one, is thus the strongest test we can make of the robustness of our approach against change of input space parametrization. For the $m_{Z^\prime}=300$~GeV signal with $N({\rm{S}})/N({\rm{R}})=10^{-3}$ and $N({\rm{R}})=2\times 10^4$, whose significance was $Z_{\rm obs}=(0.9^{+1.3}_{-0.9})\sigma$, replacing $\Delta\phi$ with $m_{ll}$ increases the observed significance to $Z_{\rm obs}=(2.3^{+1.4}_{-1.1})\sigma$.

%%%%%%%%%%%%%%%%%%%%%%%%%%%%%%%%%%%%%%%%%%%%%%%%%%%
%%%%%%%%%%%%%%%%%%%%%%%%%%%%%%%%%%%%%%%%%%%%%%%%%%%
%%%%%%%%%%%%%%%%%%%%%%%%%%%%%%%%%%%%%%%%%%%%%%%%%%%
%%%%%%%%%%%%%%%%%%%%%%%%%%%%%%%%%%%%%%%%%%%%%%%%%%%

\paragraph{Luminosity and Signal Fraction.} In the left panel of Figure~\ref{fig:lumiscale} we show our performances for the $Z^\prime$ model with $m_{Z^\prime}=300$~GeV as a function of $N({\rm{R}})$, i.e. as a function of the integrated luminosity ``$L$'' of the dataset. The observed significance shown in the plot is the median over 100 data samples with its $68\%$~C.L. error. The signal fraction is fixed to $N({\rm{S}})/N({\rm{R}})=10^{-3}$, the size of the reference sample is $\mathcal{N}_R=5 N({\rm{R}})$ and we increase $N({\rm{R}})$ from $10^4$ to $10^5$.  
Interestingly, in the regime where these tests are performed, the observed significance increases linearly with the luminosity $Z_{\rm obs}\sim L$, as opposed to the $\sqrt{L}$ growth of the reference significance. This can be explained by the fact that our analysis technique benefits from having enough statistics in the data to accurately reproduce the likelihood ratio. So increasing $L$ does not only make the signal more abundant and easier to see as in standard model-dependent analyses, but it also helps the learning process to reconstruct the most powerful (likelihood ratio) discriminant to detect it. Note however that at some point this behaviour must change and match the usual $\sqrt{L}$ scaling; this is expected to happen for very large signals corresponding to very large 
$Z_{\rm ref}$, somewhat beyond the regimes typically relevant for new physics searches.
%As anticipated we find that the observed significance increases linearly with the luminosity $Z_{\rm obs}\sim L$, as opposed to the $\sqrt{L}$ growth of the reference significance. This is not surprising since our analysis technique benefits from having enough statistics in the data to accurately reproduce the likelihood ratio. So increasing $L$ does not only make the signal more abundant and easier to see as in standard model-dependent analyses, but it also helps the learning process to reconstruct the most powerful (likelihood ratio) discriminant to detect it. Note however that at some point this behavior must saturate as the observed significance can never exceed the reference one.
%On the contrary, 

Increasing the signal fraction $N({\rm{S}})/N({\rm{R}})$ at fixed luminosity has the only benefit of increasing the ideal significance and its estimate. So both $Z_{\rm obs}$ and $Z_{\rm ref}$ increase linearly with the signal fraction as show in the right panel of Figure~\ref{fig:lumiscale}. This study was performed on the $m_{Z^\prime}=300$~GeV sample with $N({\rm{R}})=2\times 10^4$, $\mathcal{N}_R=10^5$ as for the study of the luminosity in the same figure. 

\paragraph{(In)Sensitivity to data selection.} As discussed in the Introduction, traditional model-independent strategies based on countings in bins suffer from the presence of regions in the phase space that are insensitive to new physics, because of uncorrelated Poisson fluctuation in the corresponding bins. In our approach this effect is greatly reduced, because the smoothness of the neural network protects it from following the bin-by-bin statistical fluctuations~\cite{DAgnolo:2018cun}. One particular implication of this fact is that we expect our sensitivity to depend weakly on the presence or on the absence of selection cuts that eliminate signal-free regions of the phase space. This is illustrated by studying the dependence of the observed sensitivity on: 1) a cut on the $p_T$ of the leading muon and 2) a cut on the invariant mass of the di-muon system. The $300$~GeV $Z^\prime$ model, with $\mathcal{N}_R=10^5$ and $N({\rm{S}})/N({\rm{R}})=1\times10^{-4}$ (before selection) is considered for this investigation. 

We find that the $p_T$ cut does not alter our sensitivity. For instance  the median $Z_{\rm obs}$ remains at $1\sigma$ after a $p_T>75$~GeV selection, in spite of the fact that the cut rejects $96\%$ of the background and only $5\%$ of the signal. The selection on the invariant mass instead slightly increases our sensitivity. For example $m_{ll}>95$~GeV (that rejects $90\%$ of the background and nothing of the signal) increases the median significance to $Z_{\rm obs}=1.4\sigma$.\footnote{To face the reduced amount of training data, a less complex neural network is used for this study: the 5-5-5-5-1 architecture is replaced by a 5-5-5-1. Also in this case the weight clipping and the number of training rounds are optimized following the procedure described in Section~\ref{sec:alg}.} We have observed this phenomenon already in Figures~\ref{fig:sensitivity}~and~\ref{fig:mInd}.

\begin{figure}[!t]
		\centering
	$ \begin{array}{c}
	       \includegraphics[width=0.98\linewidth]{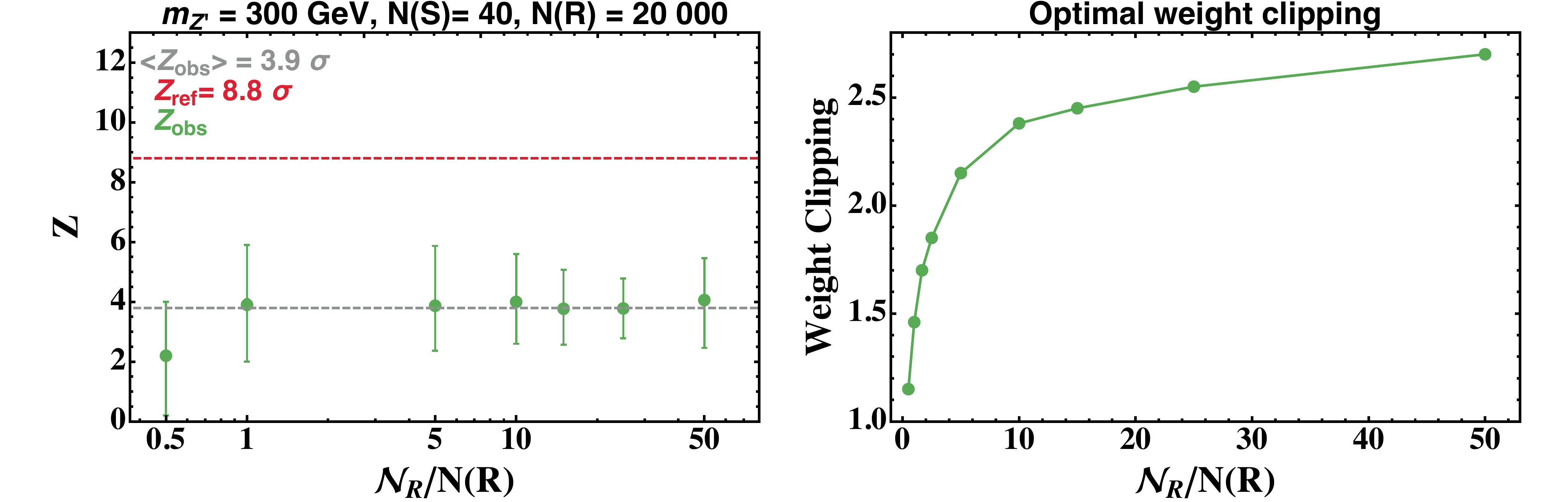}\\
	\end{array} $
	\caption{(Left) Observed significance as a function of the statistical error on the reference sample in the $m_{Z^\prime}=300$~GeV case. (Right) Optimized weight clipping as a function of reference size $\mathcal{N}_R$.}
	\label{fig:weightclipping}
\end{figure}

\paragraph{Reference Size and Optimal Weight Clipping} An accurate knowledge of known processes in the phase space of interest is crucial for the success of any new physics search. Therefore the size $\mathcal{N}_R$ of the Reference Sample should be taken as large as possible, compatibly with the computational price for training. To give an idea of the needed reference sample size, we study the performances of our method as a function of $\mathcal{N}_R/N(R)$. The result on the left panel of Figure~\ref{fig:weightclipping} is reassuring: the sensitivity is very stable as a function of this ratio up to $\mathcal{N}_R/N(R)\approx 1$. Below this value the statistical error on the reference sample in the signal region becomes sizable. If we define $\varepsilon \equiv 1/\sqrt{\mathcal{N}_R(278\leq m \leq 322)}$, i.e. counting only events in the invariant mass window populated by the signal (see Section~\ref{sec:datasets}) then the first point in the left panel of Figure~\ref{fig:weightclipping}, where $Z$ is degraded, corresponds to $\varepsilon\sim 1/2$. However this holds for a specific signal ($m_{Z^\prime}=300$~GeV $N({\rm{S}})/N({\rm{R}})=2\times10^{-3}$ and $N({\rm{R}})=2\times 10^4$), in general we expect that a degradation of the performances might be observed if $\mathcal{N}_R$ is not well above $N({\rm{R}})$, because of the result shown on the right panel of Figure~\ref{fig:weightclipping}. The plot shows the evolution with $\mathcal{N}_R/N({\rm{R}})$ of the Weight Clipping parameter, selected with the criteria of Section~\ref{sec:method}. The Weight Clipping becomes stable for $\mathcal{N}_R/N({\rm{R}})\gtrsim10$, but it abruptly drops for smaller values of this ratio. Small Weight Clipping reduces the flexibility of the neural network, which is thus less suited to identify complex new physics signals. Employing Reference samples with $\mathcal{N}_R/N({\rm{R}})\gtrsim10$, slightly above the benchmark $\mathcal{N}_R/N({\rm{R}})=5$ we employed here, is thus recommended.

\section{Conclusions and Outlook}\label{sec:conc}
We have discussed a new physics search strategy that is ``model-independent'' (i.e., not targeted to a given new physics model), with the alternative hypothesis needed for hypothesis testing being provided by a neural network. This approach was proposed in Ref.~\cite{DAgnolo:2018cun}. In this paper we made progress on its implementation and on the study of its performances. 

The main methodological advance, described in Section~\ref{sec:alg}, consists of a strategy to select the hyperparameters associated with the neural network and its training, prior to the experiment, and without relying on assumptions on the nature of the putative new physics signal. It is crucial to identify one such strategy in order to avoid the look-elsewhere effect from the ambiguities in the choice of the hyperparameters. The one we propose is heuristic, but convincing, and reduces the ensemble of hyperparameters choices to a manageable level. Progress might come on this aspect from a more sharp notion of neural network ``flexibility'' (or Capacity). Notice however that the concrete impact of the hyperparameters on the sensitivity to new physics signal has been observed to be extremely limited in all the examples we studied. Even if no systematic study has been performed, this suggests that residual ambiguities in the hyperparameters selection could be ignored.

It is not easy to quantify the performances of a model-independent search strategy. The assessment unavoidably relies on the selection of putative new physics models that are potentially present in the data, which we can try to make as broad and varied as possible. Once this choice is made, one way to proceed is to compare the sensitivity to other model-independent strategies. This is what we did in Section~\ref{sec:otherworks}, finding that our approach compares favorably to other ideas recently proposed in the literature. This comparison is however highly incomplete because it is based on a few toy problems, which are not representative of realistic LHC datasets and where new physics is extremely easy to see with our method. This is a second direction in which further work is needed.

We also need to quantify the performances in absolute terms. To this end, the most indicative quantity is arguably the probability to observe a tension with the SM if the data follow the new physics distribution. That is, the probability for our analysis to produce an interesting result. This is shown in Figure~\ref{fig:sensitivity} for different levels of observed tension and as a function of the ideal median significance of the putative new physics signal. The latter quantity, defined in Ref.~\cite{DAgnolo:2018cun} and in Section~\ref{sec:datasets}, serves as an objective measure of how ``easy-to-detect'' the new physics scenario is. Notice that the ideal significance is not the target of our method. The ideal significance can be reached, because of the Neyman--Pearson lemma, only in a fully model-dependent search where all the details of the new physics scenario are known. It cannot be obtained with any model-independent approach. With this in mind, one can still compare the observed and ideal significance directly as in Figure~\ref{fig:mInd}. The picture displays a good correlation between the ideal and observed significance in a given experiment and a weak dependence on the type of signal that is responsible for the discrepancy. This behavior might have a deep explanation, which is worth trying to identify. Yet another direction for future work is the assessment of the performances presented for more complex final states than dimuon and for more exotic putative signals. 

All the items listed above are worth investigating. However the most pressing aspect to be explored in view of the application of our strategy to real data is the inclusion of the systematic uncertainties in the reference (SM) Monte Carlo. This is conceptually straightforward because our method is based on the Maximum Likelihood approach to hypothesis testing, and systematic uncertainties are easily included in that framework as nuisance parameters. All steps needed to turn likelihood maximization into a training problem are straightforwardly repeated in the presence of nuisance parameters, as mentioned in Ref.~\cite{DAgnolo:2018cun}. The final outcome is simply that training should be performed against a reference Monte Carlo sample where the nuisance parameters are set to their best-fit values for the dataset under consideration. The concrete implementation of the algorithm in the presence of nuisance parameters thus requires two steps. The first one is to fit the nuisance parameters under the SM hypothesis to the observed data, including auxiliary measurements. Since this first step is the same as in any other experimental analysis, it should not pose any specific issue. Implementing the second step is instead problematic because it would require running the Monte Carlo with the nuisance parameters set to the observed best-fit value. Doing so for many toy SM datasets would be computationally very demanding or unfeasible. Potential solutions are either to obtain the reference sample by reweighting (which will require fitting the dependence on the nuisance of the SM likelihood possibly with a neural network) or to employ a reference sample with benchmark nuisance and correct the test statistics by some approximation of the ratio between the best-fit and the benchmark SM likelihood. It is important to verify if and how these solutions work in practice.

Before concluding it is worth outlining that the problem we are addressing is of rather general relevance in Data Analysis. The methods we are developing could thus find applications outside the specific domain of new physics searches at collider. In abstract terms, the problem can be phrased in terms of two distinct datasets, each of which can be of natural or artificial origin. The first set of data, obeying the ``Reference'' probability model, must be more abundant than the ``Data'' because it has to provide both the Reference dataset used for training and the Reference-distributed toy data used to compute the test statistic distribution. In these conditions are met, ours is a strategy to tell if the two datasets are thrown from the same statistical distribution or not, which could be useful in different domains of science. Still remaining in the context of particles physics, other potential applications of our strategies are the comparison of different Monte Carlo generators and data validation.

%%%%%%%%%%%%%%%%%%%%%%%%%%%%%%
\section*{Acknowledgments:}
%RTD is supported by the U.S. Department of Energy under Contract No. DE-AC02-76SF00515. 
MP and GG are supported by the European Research Council (ERC) under the European Union's Horizon 2020 research and
innovation program (grant agreement n$^o$ 772369). We acknowledge partial support from the Swiss National Science Foundation under contract 200021-178999. RTD acknowledges support by the Munich Institute for Astro- and Particle Physics (MIAPP) which is funded by the Deutsche Forschungsgemeinschaft (DFG, German Research Foundation) under Germany's Excellence Strategy EXC-2094-390783311. RTD acknowledges support by KITP, through the National Science Foundation under Grant No. NSF PHY-1748958. This work was partly supported by the Italian Ministry of Research (MIUR) under the PRIN grant 2017FMJFMW.

\newpage
\appendix

\section*{Appendix: Effects of reference sample's mis-modelling}
%MIS-MODELING EFFECTS (a spoiler of the future work)

For this technique to be applied to a New Physics search on real data one needs to carefully match the reference sample to the data. 
Possible sources of mis-modelling need to be studied and corrected.

The method presented in this work aims at catching discrepancies between the data and the reference, i.e. the best possible
description of the SM predictions. At this stage, no distinction is made upon the source of such discrepancies. 
Therefore a similar response is expected to effects which arise from systematic errors affecting the reference sample
%(which could be properly modelled by nuisance parameters) 
and those originated from New Physics phenomena.
In the case systematics uncertainties are not properly assessed and coped with, 
the method would likely lead to type I errors, e.g. false positives.

As already mentioned in Section \ref{sec:conc}, the method can indeed be extended to include and treat systematic 
uncertainties as nuisance parameters; the details about this procedure are being worked out and will be properly documented 
in a future publication \cite{future}. 
In this appendix we verify the aforementioned hypothesis about how a bias in the reference sample would impact the final result.

The benchmark examples addressed in sections 4 and 5 are again considered, introducing this time
an artificial mis-modelling. 
%i.e. a bias on the momentum scale of the final state's particles. 
In order to represent a realistic systematic uncertainty, a mis-calibration of the muon momentum
scale is assumed;  the relative error is typically of the order of $0.1\%$, see for instance \cite{Sirunyan:2018fpa};
still, larger values are also considered in the following, to encompass the cases with final
state objects measured with worse accuracy than muons.
%Such bias is applied
The mis-calibration is applied separately to the central ($|\eta|<1.2$) and the forward ($1.2<|\eta|<2.4$) pseudorapidity regions;
while the effects are treated as fully correlated, the magnitude in the forward region 
is assumed three times larger than in the central part.
Furthermore, as the decay of the Z boson to muons is usually exploited as a "candle" for in situ 
calibration, those events are excluded from the analysis, selecting the cases where the mass
of the lepton pair is larger than 100 GeV.

\begin{figure}[!h]
	\centering
	\includegraphics[width=0.5\linewidth]{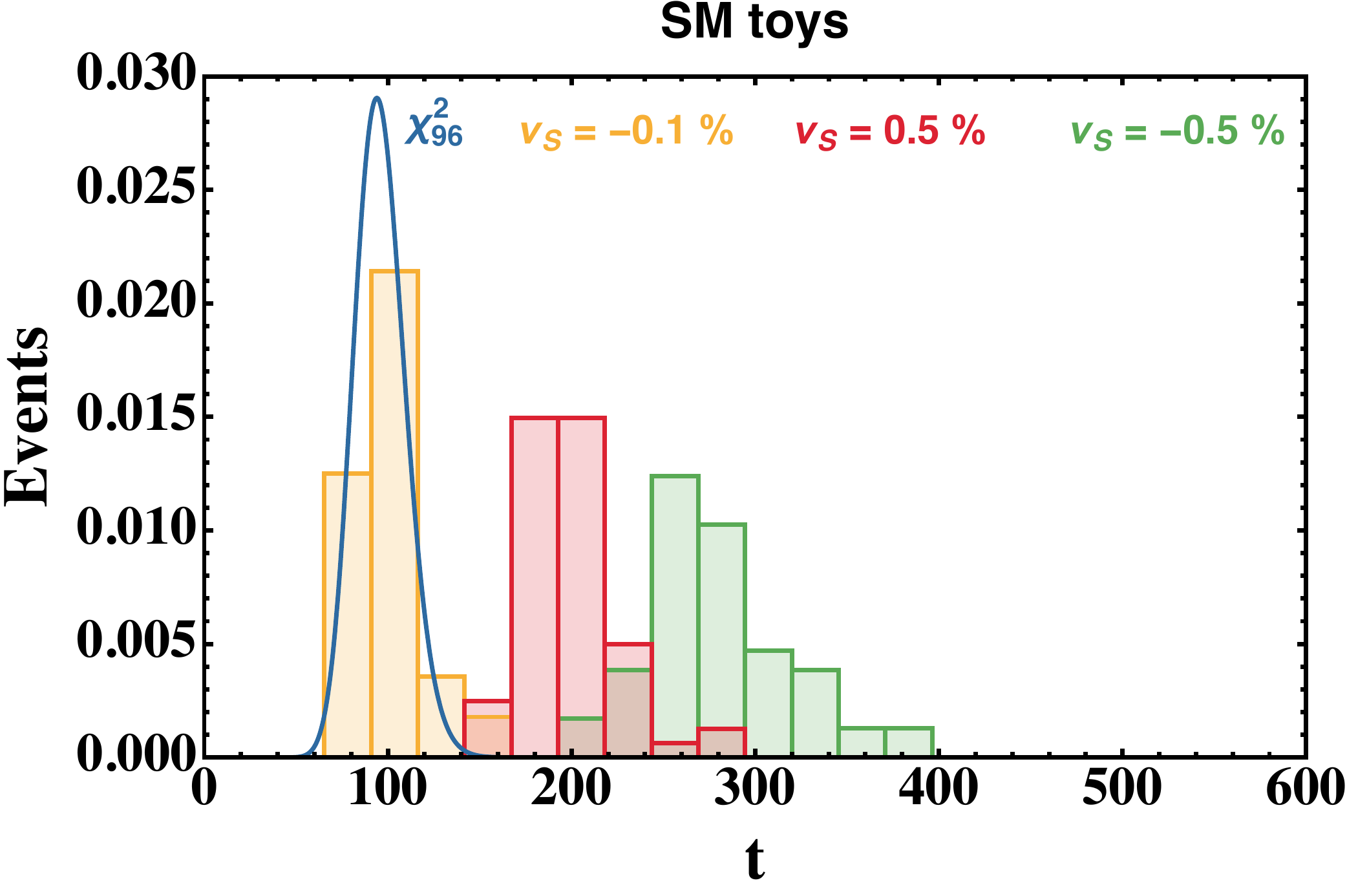} %\\ \mbox{} \\
	\caption{Test statistic distributions for toys generated accordingly to the Standard Model, in the case of $-0.1\%$ (yellow), 
	$0.5\%$ (red) and $-0.5\%$ (green) mis-calibration of the  momentum scale.}
	\label{fig:mis-modelling_SM}
\end{figure}

The response of our test statistic to artificially injected bias is checked both for toys generated according to the Standard Model 
and for toys containing New Physics. 
As far as the former are concerned, mis-calibrations of the  momentum scale of $-0.1\%$ and $\pm 0.5\%$ have been tested
(with those values referring to the central pseudorapidity region):
as can be seen from the plot in Figure \ref{fig:mis-modelling_SM}, a mis-modeling at the per mil level is not revealed by the
algorithm, whereas for larger biases higher values of the test statistics are found and thus a smaller p-value; as expected,
the method yields a false positive.

%\cite{Sirunyan:2018fpa}

%\vspace{2mm}

%\begin{figure}[!h]
%	\centering
%	\includegraphics[width=0.5\linewidth]{Figures/mis-modeling1.pdf} %\\ \mbox{} \\
%	%\includegraphics[width=0.3\linewidth]{Figures/appendix_fig1}
%	\caption{Test statistic distributions for toys generated accordingly to the Standard Model, in the case of $-0.1\%$ (yellow), 
%	$0.5\%$ (red) and $-0.5\%$ (green) mis-calibration of the  momentum scale.}
%	\label{fig:mis-modelling_SM}
%\end{figure}

\begin{figure}[!h]
	\centering
	\includegraphics[width=0.99\linewidth]{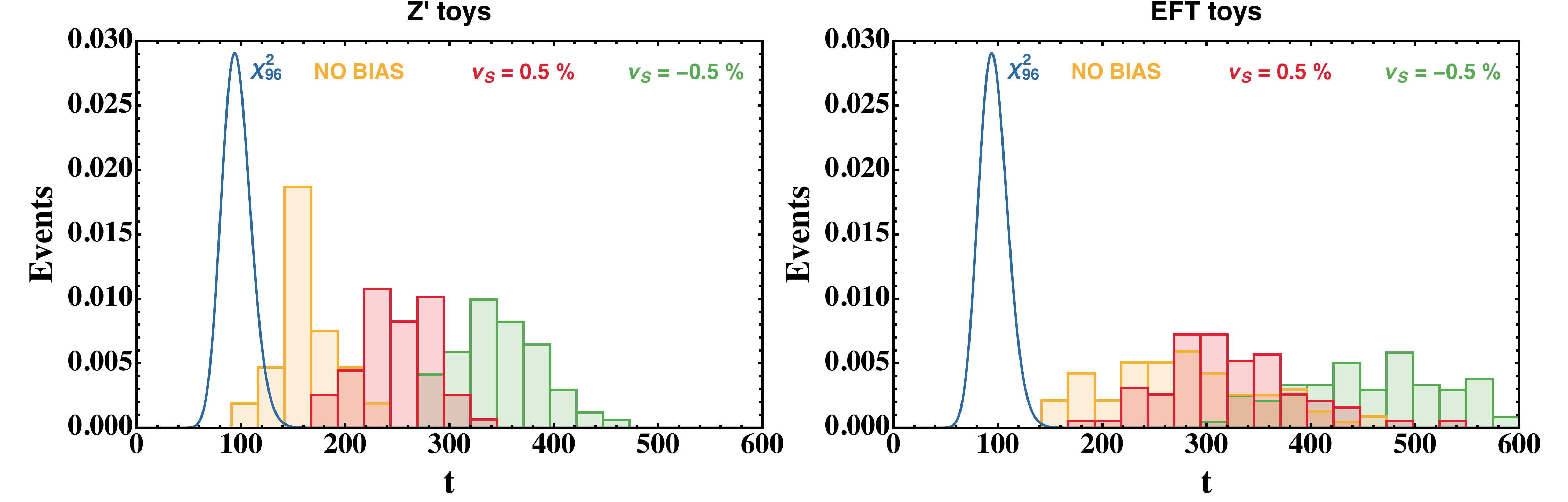} %\\ \mbox{} \\
	\caption{Test statistic distributions for toys generated with New Physics signals: $Z'$ on the left, a EFT signal on the right.
	The colours of the histograms represent respectively the cases of no mis-modelling (yellow), $0.5\%$ (red) and $-0.5\%$ (green)  
	mis-calibration of the muon momentum scale.
	}
	\label{fig:mis-modelling_NP}
\end{figure}

Testing toy datasets including New Physics effects against a mis-modelled reference sample results in the 
test statistic distributions shown in Figure \ref{fig:mis-modelling_NP}. 
We consider a $Z'$ signal (plot on the left) similar to what used for previous tests, corresponding to a reference significance of about $10\sigma$, 
%We consider a $Z'$ signal model with $m_{Z'} = 200$ GeV similar to what used for previous tests (plot on the left), corresponding to a reference significance of about $10\sigma$ and
yielding an observed significance of $4\sigma$ when tested against the unbiased reference sample.
If a mis-calibration of  $\pm 0.5\%$ on the momentum scale is introduced, the discrepancy between the reference sample and the New Physics
toys increases, leading to larger values of the test statistics.
The same happens when an EFT signal (with $c_W=10^{-6}$ TeV$^{-2}$) is injected and compared to the reference sample:
the mis-modelling of the latter enhances the significance of the signal.

In conclusion, systematic errors in the description of the SM reference sample lead to type I errors; in the case
New Physics is present in the data, the corresponding signal is not hidden by the mis-modelling 
(i.e. we do not incur into false negatives), on the contrary its significance increases.

\newpage

\bibliographystyle{utphys}
\bibliography{bibliography}

\end{document}